\documentclass[aps,
               pra,
               showpacs,
               amsmath,amssymb,
               nofootinbib,
               superscriptaddress,
               twocolumn]{revtex4-2}
\usepackage[colorlinks]{hyperref}
\usepackage[ansinew]{inputenc}
\usepackage{bbold}
\usepackage{bm}
\usepackage{amsbsy}
\usepackage{amsthm}
\usepackage{amssymb}
\usepackage{amsfonts}
\usepackage{amsmath, mathtools}
\usepackage{dsfont}
\usepackage{graphicx}
\usepackage{epsfig}
\usepackage{epstopdf}
\usepackage{dsfont}
\usepackage{mathrsfs}
\usepackage[dvipsnames]{xcolor}
\usepackage{gensymb}
\makeatletter
\newcommand\org@hypertarget{}
\let\org@hypertarget\hypertarget
\renewcommand\hypertarget[2]{%
  \Hy@raisedlink{\org@hypertarget{#1}{}}#2%
  }
\makeatother
\usepackage[figure,table]{hypcap}
\usepackage{enumerate}
\usepackage{float}
\usepackage{comment}
\usepackage{braket}
\usepackage{subfigure}
\usepackage{siunitx}
\usepackage{physics}

\hypersetup{
	bookmarksnumbered,
	pdfstartview={FitH},
	citecolor={darkgreen},
	linkcolor={darkred},
	urlcolor={darkblue},
	pdfpagemode={UseOutlines}}
\definecolor{darkgreen}{RGB}{50,190,50}
\definecolor{darkblue}{RGB}{0,0,190}
\definecolor{darkred}{RGB}{238,0,0}
\usepackage{soul}
\usepackage{orcidlink}
\usepackage{multirow} 
\usepackage[most]{tcolorbox}

\usepackage{mathtools}
\usepackage{xparse}
\makeatletter
\newcommand{\raisemath}[1]{\mathpalette{\raisem@th{#1}}}
\newcommand{\raisem@th}[3]{\raisebox{#1}{$#2#3$}}
\makeatother
\NewDocumentCommand{\newhbar}{O{0pt} O{0pt}}{
  \ensuremath{\mathrlap{\raisemath{#2}{\hspace*{#1}{\mathchar'26\mkern-9mu}}}h}%
}

\def\hbar{\newhbar[0.4pt][-0.35pt]}
\DeclareSIUnit \hb  {\textit{\hbar}}
\DeclareSIUnit{\angstrom}{\textup{\text{\AA}}}

\DeclareMathOperator{\SNR}{SNR}
\DeclareMathOperator{\CL}{CL}

\usepackage{calligra}

\DeclareMathAlphabet{\mathcalligra}{T1}{calligra}{m}{n}
\DeclareFontShape{T1}{calligra}{m}{n}{<->s*[2.2]callig15}{}
\newcommand{\scriptr}{\mathcalligra{r}\,}

\renewcommand{\vec}[1]{\boldsymbol{#1}}

\newcommand{\la}{\langle}
\newcommand{\ra}{\rangle}

\newcommand{\Order}{\mathcal{O}}

\newcommand{\da}{\dagger}

\newcommand{\Op}[1]{\hat{#1}}

\newcommand{\orho}{\hat{\rho}}
\newcommand{\osigma}{\Op{\sigma}}

\newcommand{\ovarphi}{\hat{\varphi}}

\newcommand{\oL}{\Op{L}}

\newcommand{\oa}{\Op{a}}
\newcommand{\ob}{\Op{b}}
\newcommand{\ox}{\Op{x}}
\newcommand{\oy}{\Op{y}}
\newcommand{\oz}{\Op{z}}

\newcommand{\oH}{\Op{H}}

\newcommand{\oF}{\Op{F}}

\newcommand{\oE}{\Op{E}}

\newcommand{\oK}{\Op{K}}

\newcommand{\oU}{\Op{U}}
\newcommand{\oV}{\Op{V}}

\newcommand{\oS}{\Op{S}}

\renewcommand{\op}{\Op{p}}

\newcommand{\opr}{\Op{r}}
\newcommand{\oPi}{\Op{\Pi}}

\newcommand{\vc}{\mathbf{c}}

\newcommand{\ve}{\mathbf{e}}

\newcommand{\vn}{\mathbf{n}}
\newcommand{\vp}{\mathbf{p}}

\renewcommand{\vu}{\mathbf{u}}

\newcommand{\vx}{\mathbf{x}}
\newcommand{\vy}{\mathbf{y}}

\newcommand{\vA}{\mathbf{A}}
\newcommand{\vB}{\mathbf{B}}

\newcommand{\im}{\mathrm{i}}
\newcommand{\e}{\mathrm{e}}
\newcommand{\s}{\mathrm{s}}

\newcommand{\cD}{\mathcal{D}}
\newcommand{\cE}{\mathcal{E}}
\newcommand{\cF}{\mathcal{F}}
\newcommand{\cG}{\mathcal{G}}

\newcommand{\cI}{\mathcal{I}}
\newcommand{\cJ}{\mathcal{J}}

\newcommand{\cO}{\mathcal{O}}

\newcommand{\cR}{\mathcal{R}}

\newcommand{\cT}{\mathcal{T}}
\newcommand{\cV}{\mathcal{V}}

\newcommand{\id}{\ensuremath{\mathbb{1}}}
\renewcommand{\Im}{\ensuremath{{\rm Im}}}
\renewcommand{\Re}{\ensuremath{{\rm Re}}}

\newcommand{\diff}{\mathrm{d}}

\newcommand{\vmu}{\boldsymbol{\mu}}
\newcommand{\vxi}{\boldsymbol{\xi}}

\newcommand{\ovA}{\hat{\mathbf{A}}}
\newcommand{\ovx}{\hat{\mathbf{x}}}
\newcommand{\ovp}{\hat{\mathbf{p}}}

\newcommand{\ovL}{\hat{\mathbf{L}}}
\newcommand{\ovsigma}{\hat{\boldsymbol{\sigma}}}

\renewcommand{\thesection}{\Roman{section}}
\renewcommand{\thesubsection}{\Roman{section}.\Alph{subsection}}
\renewcommand{\thesubsubsection}{\Roman{section}.\Alph{subsection}.\arabic{subsubsection}}
\makeatletter
\renewcommand{\p@subsection}{}
\renewcommand{\p@subsubsection}{}
\makeatother

\usepackage{tcolorbox}
\definecolor{c1}{HTML}{F26035} 
\colorlet{h1}{c1!30}
\definecolor{c2}{rgb}{0.0, 0.51, 0.5} 
\colorlet{h2}{c2!30}

\begin{document}
\title{Quantum Metrology of Spin Sensing with Free Space Electrons}
\author{Santiago Beltr{\'a}n-Romero\,\orcidlink{0009-0000-0310-5551}}
\thanks{These authors share first authorship
\\
santiago.romero@tuwien.ac.at\\
michael.gaida@uni-ulm.de}
\affiliation{Atominstitut, Technische Universit{\"a}t Wien, Stadionallee 2, 1020 Vienna, Austria}
\affiliation{University Service Centre for Transmission Electron Microscopy, TU Wien, Stadionallee 2, 1020 Vienna, Austria}

\author{Michael Gaida\,\orcidlink{0000-0003-3946-725X}}
\thanks{These authors share first authorship
\\
santiago.romero@tuwien.ac.at\\
michael.gaida@uni-ulm.de}
\affiliation{Institute for Complex Quantum Systems and Center for Integrated Quantum Science and Technology, Ulm University,
Albert-Einstein-Allee 11, D-89069 Ulm, Germany}

\author{Philipp Haslinger\,\orcidlink{0000-0002-2911-4787}}
\thanks{philipp.haslinger@tuwien.ac.at}
\affiliation{Atominstitut, Technische Universit{\"a}t Wien, Stadionallee 2, 1020 Vienna, Austria}
\affiliation{University Service Centre for Transmission Electron Microscopy, TU Wien, Stadionallee 2, 1020 Vienna, Austria}

\author{Dennis R{\"a}tzel\,\orcidlink{0000-0003-3452-6222}}
\thanks{These authors share last authorship\\
dennis.raetzel@tuwien.ac.at\\
stefan.nimmrichter@uni-siegen.de}
\affiliation{Atominstitut, Technische Universit{\"a}t Wien, Stadionallee 2, 1020 Vienna, Austria}
\affiliation{University Service Centre for Transmission Electron Microscopy, TU Wien, Stadionallee 2, 1020 Vienna, Austria}
\affiliation{ZARM, University of Bremen, 28359 Bremen, Germany}

\author{Stefan Nimmrichter\,\orcidlink{0000-0001-9566-3824}}
\thanks{These authors share last authorship\\
dennis.raetzel@tuwien.ac.at\\
stefan.nimmrichter@uni-siegen.de}
\affiliation{Naturwissenschaftlich-Technische Fakult{\"a}t, Universit{\"a}t Siegen, Walter-Flex-Str. 3, 57068 Siegen, Germany}

\date{\today}

\begin{abstract}

Recent advances in transmission electron microscopy (TEM) have opened the path toward spin resonance spectroscopy with single-spin sensitivity. To assess this potential, we investigate the quantum precision limits for sensing magnetic moments with free-electron probes. Using a scattering model where an electron wavepacket interacts with a localized spin, we study two metrological tasks: estimating the magnitude of the magnetic moment and discriminating the presence of a spin. The sensitivity for a given measurement setting is generally determined by the classical Fisher information, which we benchmark against the quantum bound optimized over all measurements. We find that conventional TEM imaging can saturate the quantum bound when backaction of the probe electron onto the spin state is negligible. We also find that, when backaction is relevant, one could do better by realizing a measurement of the electron's orbital angular momentum state.  These results establish the quantum limits of spin sensing in TEM and guide the development of future experiments probing individual electron spins or nanoscale ensembles of nuclear spins.

\end{abstract}

\date{\today}

\maketitle

\section{Introduction}\label{Sec:Intro}

The last two decades have brought about significant technological advances in the field of Transmission Electron Microscopy (TEM). This includes handling of samples at cryogenic temperatures~\cite{Bai2015}, aberration corrected TEM~\cite{sawada2009, Erni2009, Hawkes2009}, and  ultrafast probing~\cite{Lobastov2005, Zewail20064DUE}. Based on these breakthroughs, dynamic electron microscopy techniques have been developed leveraging atomic-level spatial resolution~\cite{ishikawa2022} and accuracy in time down to the sub-picosecond regime~\cite{Lobastov2005, Zewail20064DUE,gaida2024attosecond, baum2007attosecond}. These capabilities now allow, for example, for time-resolved studies of topological magnetic textures such as skyrmions and vortex cores and spin waves in thin (ferro)-magnetic materials~\cite{Berruto2018, daSilva2018, Moller2020, Fu2020,Liu2025CorrelatedSpins} and stroboscopic imaging of spin dynamics~\cite{Moller2020, Liu2025CorrelatedSpins}. 

The most common and precise technique to study spin dynamics is spin resonance spectroscopy (SRS). Specifically, nuclear magnetic resonance \cite{Callaghan1993Principles}, electron spin resonance \cite{Bienfait_2015}, and ferromagnetic resonance \cite{Wang_2018} have been instrumental in the development of non-invasive quantum sensing, most notably magnetic resonance imaging (MRI)~\cite{Vlaardingerbroek1996MagneticRI} with high spatial resolution \cite{lee2001MRI}. These advances led to a significant impact of SRS in a range of disciplines, including medicine \cite{Bullmore2012}, biology \cite{Borbat2001, sivelli2022nmr}, and chemistry \cite{Weil1994, Wertz2012}.

Some of us have proposed that the integration of microwave control fields with 
electron microscopy can be used to perform pump-probe protocols for SRS  
on the TEM platform (SPINEM), prospectively on the level of single spins~\cite{Haslinger_2024}.
 
First experiments have demonstrated electron SRS inside a TEM using custom-designed sample holders with integrated microwave microresonators~\cite{jaros2024electronspinresonancespectroscopy}, enabling localized in-situ detection of MW-driven spin excitations with the electron beam~\cite{jaros2025sensingspinsystemstransmission}.
In this broader context, free-space electrons can also induce backaction and coherently drive quantum systems~\cite{Ratzel2021, Gover2020free, Zhao2021, kolb2025coherentdrivingquantummodulated}, further motivating their role in quantum sensing.
The main goal here is to exploit the volumetric imaging capabilities of TEM at sub-nanometer resolution over hundreds of nanometers~\cite{Simpson2017, Seifert2020} for SRS. This calls for a detailed analysis of the fundamental precision limitations of sensing magnetic moments based on quantum estimation theory.

\begin{figure}
    \centering
    \includegraphics[width=\linewidth]{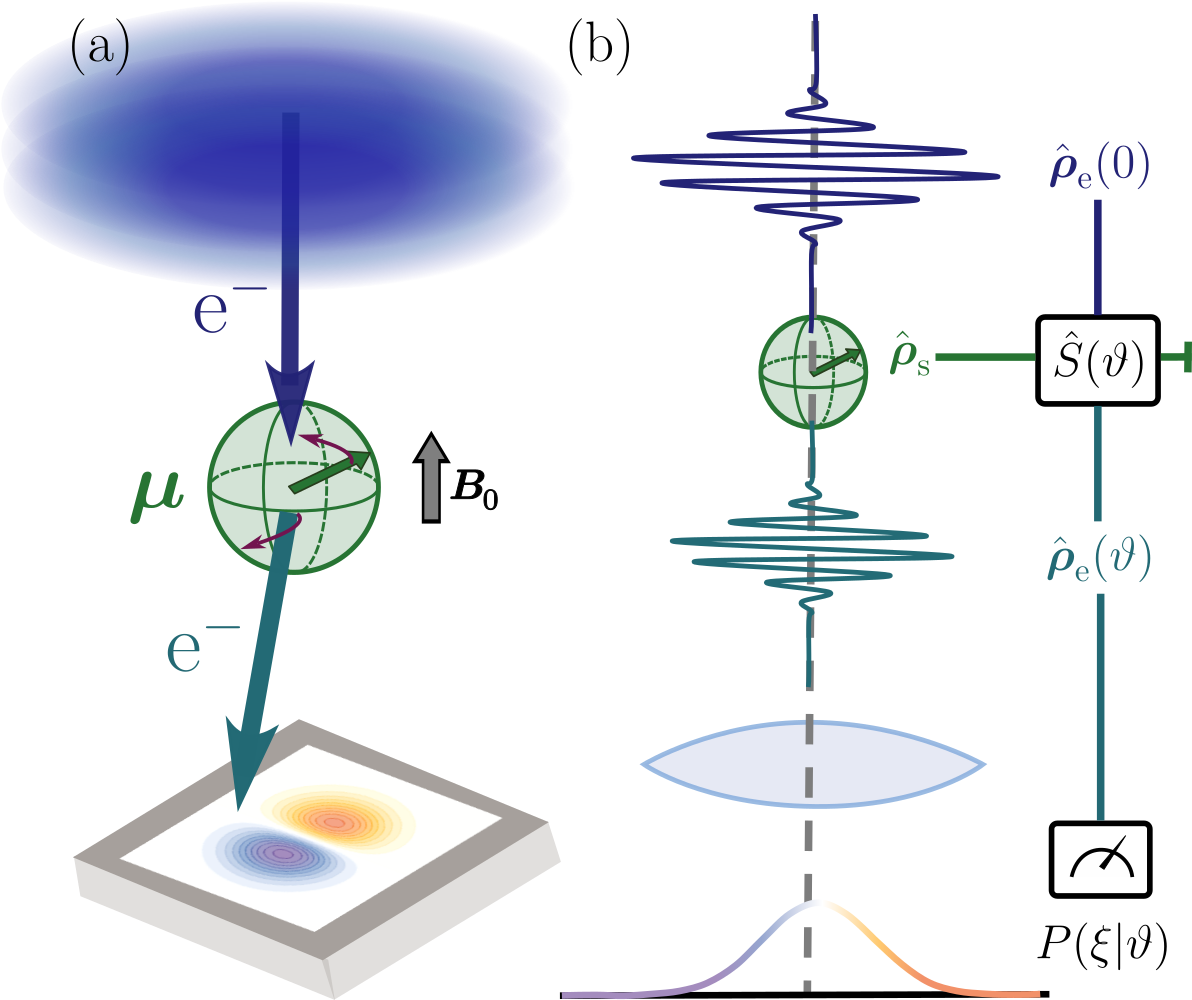}
    \caption{(a) Magnetic spin sensing scheme with a beam of free electrons. Each electron scatters off the magnetic moment $\vmu$ of a sample spin located on the beam axis in a magnetic bias field $\vB_0$, thereby exerting a backaction on the spin state (purple arrows). The electron image reveals the presence and magnitude of the magnetic moment.
    (b) We model the interaction by a scattering operator $\oS(\vartheta)$ acting on the quantum states $\orho_{\e} (0)$ and $\orho_{\s}$ of the electron and the spin, which modulates the outgoing electron state $\orho_{\e} (\vartheta)$ by a sample-dependent phase $\vartheta$. Subsequent passage through the imaging optics onto a detector screen amounts to a quantum measurement with a likelihood of outcomes $P(\xi|\vartheta)$.
}
    \label{fig:schematic}
\end{figure}

In the context of electron microscopy, methods and concepts from quantum estimation theory~\cite{nielsen2010quantum, Liu_2014, Sidhu_2020} are only recently being applied. 
Previous works have focused on the sensing of electrostatic phase shifts~\cite{Dwyer2023,Dwyer2024,evenhaim2025spinsqueezingelectronmicroscopy} and of light-induced phases~\cite{velasco2025quantumsensingmetrologyfree}. Here, we investigate how well the presence and the magnitude of a magnetic moment can be inferred from its observable effect on a passing electron in principle. We give special attention to the deep quantum regime in which the magnetic moment is sourced by a single spin in a given quantum state and no longer a fixed quantity. This highlights the impact of quantum measurement backaction and probe-sample entanglement on sensitivity, hitherto ignored, but crucial for future TEM experiments.
Our quantum metrological analysis complements another study by some of us in which a framework is developed to simulate TEM signals based on quantum electrodynamics \cite{simulationSRS_TEM}. 

The experimental scenario we consider is sketched in Fig.~\ref{fig:schematic}(a): an incoming electron wavepacket interacts with a magnetic spin localized on axis, which imprints a spin-dependent phase onto the wavepacket. The electron is then imaged onto a spatially resolving detector, and the image that forms over many repetitions encodes all accessible information about the spin. Formally, the sample interaction is described by a unitary scattering operator acting on the combined state of electron and spin, and the subsequent imaging by a unitary propagation of the electron's reduced post-scattering state, which determines the measurement basis upon detection; see Fig.~\ref{fig:schematic}(b).

Based on the acquired data, we consider two metrological tasks: (i) estimating the magnitude of the magnetic moment and (ii) discriminating wether a magnetic spin is present or not. The central figure of merit is the classical Fisher information (CFI) as it (i) quantifies the achievable estimation precision in a given setup and (ii) bounds the probability of successful discrimination in the weak-sample limit. Quantum limits of sensitivity are given by the quantum Fisher information (QFI), the maximum CFI over all possible measurement bases, and the goal is to find and realise optimal measurement schemes that saturate the QFI. 
Here, we show that conventional TEM imaging can saturate the quantum limit when backaction is negligible. With backaction, the precision improves by detecting the electron's orbital angular momentum (OAM); sorting of OAM was recently demonstrated experimentally~\cite{tavabi2021OAMsorter}. Contrary to TEM imaging, such a measurement is also sensitive to spins aligned with the beam axis. For discrimination (ii), TEM imaging is close to optimal for spins aligned perpendicular to the beam axis, whereas angular momentum measurements should only be used for spins aligned with the beam axis.

This article is organized as follows. We first introduce the scattering model for magnetic moment sensing with free electrons in Sec.~\ref{Sec:Hamiltonian}. Then we study the metrological tasks of magnetic moment estimation and discrimination in Sec.~\ref{Sec:Sensing} and \ref{Sub:Basics:StateDiscrimination}, respectively. In Sec.~\ref{Sec:Discussion}, we conclude with a discussion of our findings, realistic case studies, and an outlook.

\section{Interaction model}
\label{Sec:Hamiltonian}

We begin by modeling the physical interaction between a relativistic free probe electron and a magnetic moment sourced by a localized single spin or large spin ensemble. It will lead us to a scattering operator in the paraxial regime of high kinetic energy, which will serve as the starting point for our metrological analysis in Sec.~\ref{Sec:Sensing}. We shall omit any other contributions to the scattering signal that would be present (or even dominant) in realistic samples, such as electric dipoles. This is justified since, in an SRS experiment, one would perform differential measurements to single out a change in the scattering signal of magnetic moments, e.g., by exciting spin precession with a resonant pump pulse or by switching the externally applied magnetic bias field~\cite{Haslinger_2024,simulationSRS_TEM, jaros2025sensingspinsystemstransmission}.

Consider the minimal coupling Hamiltonian of a relativistic point charge $e$ of mass $m$ subject to the electromagnetic potentials $(\Phi(\vx), \vA(\vx))$ of a magnetic moment $\vmu$ aligned relative to a bias field $\vB_0 = B_0 \vn_0$,
\begin{equation}
    \label{eq:Met:MinCoup}
     \oH = \sqrt{m^2c^4+c^2\left( \ovp + e\vA(\ovx) \right)^2} - e \Phi(\ovx) -\vB_0 \cdot \vmu.
\end{equation}
Here, $c$ denotes the speed of light and $(\ovx,\ovp)$ are the position and momentum operators of the charge, which represents our free electron. We shall operate in the Coulomb gauge and set $\Phi \equiv 0$. A single magnetic point dipole (spin) at the origin would contribute the vector potential $\vA_0 (\ovx) = \mu_0 \vmu\times \ovx / 4\pi |\ovx|^3$, with $\mu_0$ the vacuum permeability. This singular potential would result in an unphysical divergence of sensitivity and should therefore be regularized. To this end, we introduce a finite spatial density distribution for the spin, $P_{\rm s} (\vx) \geq 0$ with $\int\diff^3\vy \, P_{\rm s}(\vy)=1$, and take the average of the vector potential as $\vA_{\rm s} (\vx) = \int\diff^3\vu \, P_{\rm s} (\vx-\vu) \vA_0 (\vu)$. For a radially symmetric distribution, the regularized potential simplifies to $\vA_{\rm s} (\vx) = F(|\vx|)\vA_0 (\vx)$, where $F$ is a smooth radial function that alleviates the singularity at the origin, 
\begin{equation}\label{eq:F}
   F(x) = 2\pi x^2 \int_0^\infty \diff u \int_{-1}^1 \diff \tau \, \tau \, P_\mathrm{s}\left( \sqrt{u^2 + x^2 - 2u x \tau} \right); 
\end{equation}
see App.~\ref{App:HamDerivation} for a derivation. Denoting the characteristic width of this function by $\Delta_{\rm s}$, the maximum magnitude of the vector potential is thus roughly given by $A_{\max} \sim \mu_0 |\vmu| /\Delta_{\rm s}^2$.

In the relevant regime of high kinetic energies (of the order of $10^2\,$keV in TEM), we can treat the magnetic potential as a small perturbation and expand the Hamiltonian \eqref{eq:Met:MinCoup} to lowest order ,
\begin{subequations}\label{eq:Met:RegHam}
    \begin{align}
\oH =&  \sqrt{m^2c^4+c^2\left| \ovp\right|^2} -\vB_0 \cdot \vmu  +\oV, \label{eq:H0}\\
    \oV:=& \frac{\hbar r_\mathrm{e}}{2m} \left( \frac{\vmu}{\mu_\mathrm{B}} \right) \cdot \left( \frac{F(|\ovx|)}{|\ovx|^3} \frac{\ovx\times \ovp}{\gamma(|\ovp|)} \right). \label{eq:V}
    \end{align}
\end{subequations}
Here, $r_\mathrm{e} = \mu_0 e^2/4 \pi m \approx 2.8\cdot 10^{-15}$~m denotes the classical electron radius, $\mu_\mathrm{B} = e\hbar/2m$ the Bohr magneton, and $\gamma(p)=\sqrt{1+(p/mc)^2}$ the Lorentz factor. 
Notice that we have neglected the $\vA^2$-term in the expansion, because it is typically several orders of magnitude smaller than the leading-order interaction term, viz.~$eA_{\max} \ll m\gamma v_0$ given the mean electron velocity $v_0$.
Moreover, we ignore the probe electron's spin, because its magnetic dipole interaction energy with the sample is of the order of $A_{\max}\mu_{\rm B}/\Delta_{\rm s}$ and thus negligible as long as $m v_0 \Delta_{\rm s} \gg \hbar$. The effect of a longitudinal bias field $B_0$ on the electron beam can be safely ignored, since its deflection is either compensated by the electron optics (lenses) or eliminated in practice by image subtraction, where constant contributions not related to spin dynamics are removed in the differential analysis~\cite{simulationSRS_TEM}.

The net effect of the interaction on the probe electron (and the sample spin) is described by the scattering operator: $\oS = \cT\exp [-\im \int_{-\infty}^{+\infty} \diff t \, \oV^{(\mathrm{I})} (t)/\hbar]$,
where $\oV^{(\mathrm{I})}(t)$ is the interaction term \eqref{eq:V} in the interaction picture and $\cT$ denotes the time-ordering operator. In general, $\oS$ is not analytically tractable, but an approximate expression exists in the common TEM regime of paraxial beams of high energy and comparably weak samples~\cite{simulationSRS_TEM}. In this regime, the electron state is governed by a large central momentum pointing along the beam axis (here: $z$-axis), $\vp_0 = p_0 \ve_z$, and a wave function $|\psi\ra$ that extends over small transverse momenta, $|\vp_{\perp}| \ll p_0$. 

As detailed in App.~\ref{App:Met:ScattSVEA}, we can now expand the Hamiltonian \eqref{eq:Met:RegHam} to lowest order in the small momentum deviations from $\vp_0$ and integrate the resulting time evolution in the interaction picture. There, the sample magnetic moment will be time-dependent, $\vmu^{\rm (I)}(t)$, as it precesses about the bias field $\vB_0$ in \eqref{eq:H0}. The precession rate is however much slower than the effective interaction time with the passing electron wave packet, which amounts to femtoseconds for velocities $v_0 \sim 10^{-1}c$ and longitudinal coherence lengths $\sim 10^{-7}$m~\cite{Reimer_2008}. 
We can thus assume that the sample is probed stroboscopically and the electron sees a constant magnetic moment $\vmu$ as it passes. We arrive at the scattering operator
\begin{equation}\label{eq:S}
    \oS (\vartheta) = \exp \left[ -\im \vartheta g(\opr) \frac{\mu_x \sin \ovarphi - \mu_y \cos\ovarphi}{|\vmu|}\right],
\end{equation}
which modulates the phase of the electron wavefunction by the magnitude 
\begin{equation} \label{eq:phase}
    \vartheta = \frac{r_\mathrm{e} |\vmu|}{\mu_\mathrm{B} \Delta_{\s} }= \frac{\mu_0 e}{2\pi\hbar} \frac{|\vmu|}{\Delta_{\s}} .
\end{equation}
Here we have introduced the polar representation of the transverse coordinates, $(\ox, \oy)=\opr(\cos\ovarphi, \sin\ovarphi)$, and the radial interaction profile,
\begin{equation}\label{eq:g}
    g(r) = \frac{\Delta_{\rm s} r}{2} \int_{-\infty}^\infty \diff z \frac{F(\sqrt{r^2+z^2})}{(r^2+z^2)^{3/2}} .
\end{equation}
Notice that the scattering operator is insensitive to the magnetic moment component $\mu_z$ along the beam axis. The Lorentz factor no longer appears as the leading relativistic corrections of mass and passage time cancel.

The radial function $g$ is given via \eqref{eq:F} by a triple integral over the spatial distribution $P_{\s}$, which must be evaluated numerically in most cases. Gaussian distributions are a convenient exception; identifying $\Delta_{\rm s}$ with their standard deviation, they result in 
\begin{equation}\label{eq:g_Gauss}
    g \left(r\right) = \left(1-\e^{-r^2/2\Delta_{\s}^2}\right) \frac{\Delta_{\s}}{r}.
\end{equation}
We shall employ this simple Gaussian model for our sensitivity case studies; it matches the predictions for other, more realistic distributions reasonably well, see Fig.~\ref{fig:realistic_NeNshots} below.
As is common in TEM, we will also use an initially Gaussian transverse wavefunction of the probe electron,
\begin{equation}\label{eq:psi0}
    \la \vx_{\perp}|\psi(0) \ra = \frac{1}{\sqrt{2\pi \Delta_{\e}^2}} \e^{-r^2/4\Delta_{\e}^2},
\end{equation}
with standard deviation $\Delta_{\e}$. This focus size $\Delta_{\e}$ also determines whether the paraxial scattering operator \eqref{eq:S} is valid. A detailed assessment in App.~\ref{App:Comparison} for a typical beam energy of $200\,$keV shows that deviations are negligible if $\Delta_{\rm e} \gg 10^{-12}\,$m.

In the following metrological study, we will juxtapose two opposing scenarios: the no-backaction case (NB) in which the sample comprises a sufficiently large quasi-classical magnetic moment that is unaffected by the probe electron, and the backaction case (BA) in which the magnetic moment is sourced by a single quantum spin that undergoes state transitions due to the electron. 
The NB-scenario includes conventional magnetic samples such as ferromagnets, paramagnets, organic radicals, or ensembles of nitrogen vacancy centers. The magnetic moment arises from the collective state of many aligned atomic spins, to which the short presence of an electron is insignificant. We may thus treat the magnetic moment in \eqref{eq:S} as a fixed vector, $\vmu = \mu \vn$, and  take its transverse component as, say, our $x$-axis, $\vn \equiv |\vn_{\perp}|\ve_x + n_z \ve_z$, so that
\begin{equation}
    \label{eq:Met:ScattNoBa}
    \oS_{\rm NB}(\vartheta) = \e^{-\im \vartheta |\vn_\perp| g(\opr) \sin\ovarphi} = \sum_{k \in \mathbb{Z}} (-1)^k J_k\left[ \vartheta |\vn_\perp| g(\opr) \right] \e^{\im k \ovarphi}.  
\end{equation}
Here, we made use of the Jacobi-Anger expansion~\cite{korenev2002bessel} in terms of Bessel functions, which reveals the effect of the scattering operator on the electron's orbital angular momentum (OAM), $\ovL = \ovx \times \ovp$: each summand displaces $\oL_z$ by $k\hbar$. The electron state remains pure and transforms as $|\psi(\vartheta)\ra = \oS_{\rm NB}(\vartheta)|\psi(0)\ra$.

For the deeply quantum BA regime, consider a well-isolated spin of, say, an unpaired electron in an organic molecule or a quantum dot, or of an atomic nucleus in a biological sample. Within nanometer proximity, a probe electron can interact sufficiently strongly for spin-dependent scattering, backaction, and even measurable entanglement to occur \cite{evenhaim2025spinsqueezingelectronmicroscopy}. In this case, the magnetic moment in \eqref{eq:S} must be replaced by an operator acting on a spin-$1/2$ system, $\mu = \mu \ovsigma$, with $\ovsigma$ the vector of Pauli matrices. The scattering operator becomes
\begin{equation}\label{eq:Met:ScattBA}
    \oS_{\rm BA}(\vartheta) = \oK_0(\vartheta) 
    + \oK_-(\vartheta)\, \hat{\sigma}_+ - \oK_+(\vartheta)\, \hat{\sigma}_-,
\end{equation} 
where $\osigma_{\pm} = (\osigma_x \pm \im\osigma_y)/2$ and 
\begin{equation}\label{eq:Kops}
    \oK_0(\vartheta) = \cos[\vartheta g(\opr)], \quad 
    \oK_\pm(\vartheta) = \sin[\vartheta g(\opr)]\,\e^{\pm\im \ovarphi}.
\end{equation}
By virtue of angular momentum conservation, the sample spin imparts at most $\pm \hbar$ of OAM displacement. 
Given an arbitrary initial spin state in the Bloch vector representation, $\orho_{\s}=\left(\mathds{1}+\vec{c}\cdot\ovsigma \right)/2$ with $|\vec{c}| \leq 1$, we can once again choose our coordinates such that $\vc = |\vc_{\perp}|\ve_x + c_z \ve_z$. The initially pure electron state transforms into a mixed, partially decohered state as it gets correlated or even entangled with the sample, 
\begin{align}\label{eq:finalstate_BA}
    \orho_{\e}(\vartheta) 
    =& \tr_{\s}\left\{ \hat{S}_{\mathrm{BA}}(\vartheta) \left( |\psi(0)\ra\la \psi(0)| \otimes \orho_{\s} \right) \hat{S}_{\mathrm{BA}}^\dagger(\vartheta) \right\} \\
    =& \quad \oK_0(\vartheta) \, |\psi(0)\ra\la \psi(0)| \, \oK_0^\dagger(\vartheta) \nonumber \\
    & + \sum_{j=\pm}\frac{1+jc_z}{2} \oK_j(\vartheta) |\psi(0)\ra\la \psi(0)| \oK_j^\dagger(\vartheta) \nonumber \\
    & + i |\vec{c}_\perp| \left[ \oK_0(\vartheta) |\psi(0)\ra\la \psi(0)| \sin [\vartheta g(\opr) ] \sin\hat{\varphi} - h.c. \right]; \nonumber
\end{align}
see App.~\ref{App:Met:ScattBa} for a derivation.

\section{Magnetic moment estimation}\label{Sec:Sensing}

The interaction model at hand, we proceed with our first metrological task: estimating the magnitude of the magnetic moment from an observed electron image. It enters the scattering operator \eqref{eq:S} through the parameter $\vartheta$ and modulates the phase of the electron wavefunction in position representation. We are thus tasked with a phase estimation problem, the goal of which is to infer a given unknown $\vartheta$-value from random measurement outcomes $\xi$ that are distributed according to a known likelihood function $P(\xi|\vartheta)$.

Here, the likelihood is given by a quantum measurement on the scattered electron state $\orho_{\e}(\vartheta)$, formally described by a positive operator-valued measure (POVM), $\cE = [\oE(\xi)]_{\xi}$: a family of positive (or even projective) operators $\oE(\xi)$ that fulfill $\int\diff\xi\, \oE(\xi) = \id$ and determine the likelihood of each measurement outcome $\xi$ as
\begin{equation}
    P(\xi|\vartheta) = \tr \left\{ \orho_{\e}(\vartheta) \oE(\xi) \right\}.
\end{equation}
In TEM, the outcomes are two-dimensional coordinates of electrons on a detector plane. Depending on the mode of operation, these correspond to measurements of transverse position (image mode, $\oE(\xi) \equiv |\vx_{\perp}\ra\la \vx_{\perp}|$), momentum (diffraction mode, $\oE(\xi) \equiv |\vp_{\perp}\ra\la \vp_{\perp}|$), or linear combinations of position and momentum (out-of-focus detector). 
Although realistic detectors are limited in size and resolution, restricting the outcomes to a finite number of discrete pixel values, we will not consider such technical constraints as we are concerned with sensitivity limits of adequate detectors.
As an alternative to standard TEM detection, we will also consider a direct measurement of the electron's OAM, motivated by the fact that the scattering operator describes OAM transfer between the sample and the electron.

Consider an experiment recording $N$ independent single-electron shots with outcomes $\vxi = (\xi_1,\ldots,\xi_N)$ at a fixed unknown true $\vartheta$-value. Based on the likelihood, $P_N(\vxi|\vartheta) = \prod_n P(\xi_n|\vartheta)$, one constructs an estimator $\tilde{\vartheta}$ that assigns a parameter estimate $\tilde{\vartheta} (\vxi)$ to the recorded data. A suitable estimator should be unbiased, at least asymptotically for $N\gg 1$; i.e., its expectation value over all possible outcomes should coincide with the true value,
\begin{equation}
    \mathbb{E}_\vartheta \left[ \tilde{\vartheta} \right] = \int \diff \vxi \ \tilde{\vartheta}(\vxi) P(\vxi \vert \vartheta) = \vartheta.
\end{equation}
The variance of the estimate then equals the mean-square deviation between the estimate and the unknown true value, $\mathbb{V}_\vartheta[\tilde{\vartheta}] = \mathbb{E}_\vartheta [ ( \tilde{\vartheta} - \vartheta )^2 ]$, which quantifies the average estimation error and satisfies the Cram\'{e}r-Rao bound \cite{cramer1946,rao1945}. It can be stated as an upper bound on the signal-to-noise ratio (SNR), 
\begin{equation}
    \label{eq:CramerRaoBound}
 \SNR (\vartheta) = \frac{\vartheta}{\sqrt{\mathbb{E}_\vartheta [ ( \tilde{\vartheta} - \vartheta )^2 ]}} \leq \vartheta \sqrt{N \cI(\vartheta)}.
\end{equation}
For example, the commonly chosen maximum-likelihood estimator, $\tilde{\vartheta}(\vxi) = \arg \max_{\vartheta} P(\vxi|\vartheta)$, satisfies and even saturates this bound asymptotically \cite{fisher1925estimation}.
Here, the $\sqrt{N}$ scaling represents the shot noise limit of estimation precision for $N$ detection events. The 'classical' Fisher information (CFI) $\cI(\vartheta)$ measures the single-shot sensitivity to variations of the parameter around the value $\vartheta$,
\begin{equation} \label{eq:CFI}
    \cI(\vartheta) = \int \diff \xi \ \frac{1}{P(\xi \vert \vartheta)} \left( \frac{\partial P(\xi \vert \vartheta)}{\partial \vartheta} \right)^2.
\end{equation}
Since the true $\vartheta$-value is unknown, one evaluates the CFI for an anticipated range of values; it may even be constant. We are mostly concerned with weak samples and evaluate $\cI(\vartheta \to 0)$.

The CFI provides a benchmark for different measurement protocols in terms of their sensitivity to the magnetic phase $\vartheta$. In particular, one obtains the optimal sensitivity by maximizing the CFI over all possible POVMs \cite{braunstein_statistical_1994}, $\cJ(\vartheta) = \max_{\cE} \cI(\vartheta)$. This maximum, called the quantum Fisher information (QFI), describes the quantum limit of sensitivity of the scattered electron state $\orho_{\e}(\vartheta)$ to variations of $\vartheta$. It can be expressed in terms of the solution to a Lyapunov equation \cite{helstrom1967minimum},
\begin{equation}\label{eq:Def:QFI}
  \cJ(\vartheta) = \tr \{ \orho_{\e}(\vartheta) \oL^2 (\vartheta) \}, \ \ \text{with}\ \ \orho_{\e}' (\vartheta)   = \frac{\{ \orho_{\e} (\vartheta), \oL(\vartheta) \}}{2}  . 
\end{equation}
Here, $\orho_{\e}' = \diff \orho_{\e}/ \diff \vartheta$. The hermitean solution $\oL(\vartheta)$, known as the symmetric logarithmic derivative (SLD), is generally difficult to compute for infinite-dimensional states. For pure states, $\orho_{\e}(\vartheta) = |\psi(\vartheta)\ra\la \psi(\vartheta)|$, the SLD is simply $\oL (\vartheta) = 2\orho_{\e}' (\vartheta)$ and the QFI reduces to
\begin{equation}\label{eq:QFI_pure}
    \cJ(\vartheta) = 4 \left[ \la \psi'(\vartheta) | \psi'(\vartheta)\ra - \left| \la \psi(\vartheta)|\psi'(\vartheta)\ra \right|^2 \right],
\end{equation}
where $|\psi'(\vartheta)\ra = \diff |\psi(\vartheta)\ra/\diff \vartheta$. The eigenvectors of the SLD define a measurement basis for which the CFI saturates the QFI, allowing one to approach the optimal SNR for $N$ independent probes, $\SNR(\vartheta) \leq \vartheta \sqrt{N \cJ(\vartheta)}$. However, this optimal measurement may not always be feasible. A collective measurement on an entangled state of $N$ probes (instead of $N$ single-probe measurements) can beat the shot-noise limit and approach the so-called Heisenberg scaling, $\cJ (\vartheta) \propto N^2$ \cite{Lee2002,evenhaim2025spinsqueezingelectronmicroscopy}.

We proceed to calculate the QFI with and without backaction and compare it to the CFI for diffraction-mode and OAM measurements. Image-mode electron measurements in are insensitive in our case, because the scattering operator \eqref{eq:S} is diagonal in position representation. It is straightforward to check that the respective CFI vanishes.

\subsection{Estimation without backaction}\label{Sec:Sensing_NB}

In the NB-case, the unitary scattering transformation according to \eqref{eq:Met:ScattNoBa} preserves the purity of the electron's transverse wavefunction. Given that this wavefunction is radially symmetric, i.e., depends only on $r=|\vx_{\perp}|$, the pure-state QFI expression \eqref{eq:QFI_pure} then yields a constant sensitivity of the probe electron to any $\vartheta$-value,
\begin{equation}
    \label{eq:QFI:NB}
    \cJ_\mathrm{NB} =  |\vn_\perp|^2 \cG, \ \ \text{with}\ \ \cG =  2\bra{\psi(0)} g^2(\opr) \ket{\psi(0)}.
\end{equation}
We also see that the electron can only sense the magnitude of the magnetic moment component $\vn_{\perp}$ in the sample plane perpendicular to the beam axis. The in-plane QFI value $\cG$ further depends on the shape of the electron wavefunction and the spatial distribution of the magnetic moment; it can be brought to a simple instructive form in the Gaussian model based on \eqref{eq:g_Gauss} and \eqref{eq:psi0},
\begin{equation}
    \label{eq:cG:Gaussian}
    \cG = \frac{2}{\chi^2} \ln \left( \frac{1+\chi^2}{\sqrt{1+2 \chi^2}} \right),
\end{equation}
which depends merely on the ratio of standard deviations, $\chi = \Delta_{\e}/\Delta_{\s}$. The maximum, $\cG \approx 0.3$, is reached at $\chi \approx 1.2$. Hence, one achieves the best sensitivity by focusing the electron to no less than about the size of the magnetic moment. Sensitivity drops otherwise, when only a fraction of the wave function interacts with the magnetic moment ($\chi \gg 1$), or vice versa ($\chi \ll 1$). 

We find that the quantum limit of sensitivity is readily attainable in a conventional diffraction-mode measurement. This is due to the fact that the momentum-basis CFI saturates the QFI when the post-scattering momentum amplitudes are real-valued \cite{wasak_optimal_2016},
\begin{equation}
    \bra{\vp_\perp} \oS_\mathrm{NB}(\vartheta) \ket{\psi(0)} \in \mathbb{R}, \quad \forall \vartheta, \vp_\perp.
\end{equation}
We discuss this in App.~\ref{App:QFI} and show that in practice, the saturation requires that the detector cover momenta $|\vp_{\perp}|> \hbar/\Delta_{\s}$ and resolve $|\vp_{\perp}|< \hbar/\Delta_{\e}$. 
Imaging the electron on a defocus plane in between image mode ($\cI_{\rm NB}^{\rm (x)} = 0$) and diffraction mode ($\cI_{\rm NB}^{\rm (p)} = \cJ_{\rm NB}$) can only result in a lower estimation precision.

The optimality of momentum measurements is not apparent from the explicit form or the SLD operator. At the same time, it is striking that the attainable QFI for the Gaussian model amounts to less than a bit in phase resolution ($\cJ_{\rm NB} < 0.3$) at any $\vartheta$-magnitude, despite the sheer capacity of the possible measurement outcomes to encode many bits of information. We shall illustrate this further by constructing a binary measurement that yields approximately the same estimation precision per electron as the highly resolving momentum measurement in the limit $\vartheta \to 0$. 

Recall that the scattering operator \eqref{eq:Met:ScattNoBa} maps a radially symmetric zero-OAM state of the electron to a superposition of OAM states. Small variations of $\vartheta \neq 0$ translate into small populations of nonzero OAM components. Consider therefore the binary projective POVM, $\cE = [ \oPi, \mathds{1}-\oPi ]$, which tests whether or not the electron has zero OAM, $\oPi = \int_0^\infty \diff r \ r \ket{r,0} \bra{r,0}$, given the improper zero-OAM and radius eigenstates, $\la \vx_{\perp} | r,0\ra = \delta(|\vx_{\perp}|-r)/\sqrt{2\pi}r$. The zero-OAM likelihood, 
\begin{equation}\label{eq:zeroOAM_P}
    P_0(\vartheta) = \la \psi(\vartheta)|\oPi|\psi(\vartheta)\ra = \la \psi(0)|J_0^2 [\vartheta |\vn_{\perp}|g(\opr)]|\psi(0)\ra,
\end{equation} 
determines the CFI as
\begin{equation}\label{eq:CFI_OAM}
    \cI_\mathrm{NB}^{\rm (L)}(\vartheta) = \frac{[P_0' (\vartheta)]^2}{P_0(\vartheta)[1-P_0(\vartheta)]} =  \cJ_{\rm NB} - \cO(\vartheta^2),
\end{equation} 
which follows from a Taylor expansion of the Bessel function $J_0$.
A binary OAM measurement is thus also optimal for small $\vartheta$, in analogy to a seminal result for the imaging resolution of incoherent point sources~\cite{tsang2016quantum}.

\subsection{Estimation with backaction}\label{Sec:Sensing_BA} 

In the BA-case, we are to estimate the magnetic moment of a single (electronic or nuclear) quantum spin, which we can safely describe in the weak-sample limit, $\vartheta \to 0$. However, this comes with a non-negligible electron-spin correlations induced by the scattering operator \eqref{eq:Met:ScattBA}, resulting in the non-unitary scattering transformation \eqref{eq:finalstate_BA} of the reduced electron state. A general expression for the QFI of this mixed scattered state $\orho_{\e} (\vartheta)$ is no longer available, but for radially symmetric initial states $|\psi(0)\ra$ and small $\vartheta$, we can invoke upper and lower bounds to show that
\begin{equation}
    \label{eq:QFIBA}
    \cJ_\mathrm{BA} = 2 \cG - \cO\left( \vartheta^2 \right).
\end{equation}
First, to prove the upper bound $\cJ_{\rm BA} \leq 2\cG$, let the initial spin state also be pure, $\orho_{\s} = |\psi_{\s} \ra \la \psi_{\s}|$. The unitary scattering operator \eqref{eq:Met:ScattBA} then maps the initial product state to an entangled electron-spin state, whose QFI can be obtained through \eqref{eq:QFI_pure},
\begin{align}
    \label{eq:PureQFIBa}
    \cJ_\mathrm{tot}
    &=  4 \bra{\psi(0),\psi_{\s}} g^2(\opr) \underbrace{\left[ \sin(\hat{\varphi}) \hat{\sigma}_x - \cos(\hat{\varphi}) \hat{\sigma}_y  \right]^2}_{=\mathds{1}} \ket{\psi(0),\psi_{\s}} \nonumber \\ 
    &= 4 \bra{\psi (0)} g^2(\opr) \ket{\psi (0)} = 2 \cG.
\end{align}
This is twice the maximum QFI in the NB-case, regardless of the initial spin state. Hence, the QFI cannot increase by using a mixed spin state instead. Neither does it increase by averaging out the spin degree of freedom \cite{petz1994geometry}, which results in the reduced electron state. Therefore, $\cJ_{\rm BA} \leq \cJ_\mathrm{tot} = 2\cG$.

Lower bounds to the QFI are prodiced by the CFI of any measurement on the scattered electron state \eqref{eq:finalstate_BA}. Once again, the binary OAM measurement turns out to be optimal. For the zero-OAM likelihood, we obtain
\begin{equation}\label{eq:zeroOAM_PBA}
     P_0(\vartheta) = \bra{\psi(0)} \cos^2 [\vartheta g(\opr) ] \ket{\psi(0)} = 1 - \frac{\vartheta^2}{2} \cG + \cO \left( \vartheta^4 \right).
\end{equation}
Remarkably, the corresponding CFI is twice the CFI (and QFI) that could at most be achieved in the NB-case,
\begin{equation}
    \cI_\mathrm{BA}^{\rm (L)}(\vartheta) = 2 \cG - \cO \left( \vartheta^2 \right).
\end{equation}
Thanks to the backaction, the OAM state of the electron contains more, not less, information about the sample spin. Since $\cI_\mathrm{BA}^{\rm (L)} \leq \cJ_\mathrm{BA}$, we accomplish our proof of \eqref{eq:QFIBA}.

Contrary to the NB-case, diffraction mode imaging is no longer optimal. In fact, we show in App.~\ref{App:QFI} that
\begin{equation}
    \cI_\mathrm{BA}^{(\mathrm{p})}(0) = |\vc_\perp|^2 \cG.
\end{equation}
In other words, the diffraction-mode CFI with backaction matches the one without backaction for $\vartheta\to 0$, only with the Bloch vector component $\vc_{\perp}$ of the spin state in place of $\vn_{\perp}$ before. The backaction effect does not reduce the estimation precision in diffraction mode, but instead doubles the quantum-limited precision for an OAM measurement.

\begin{table}
\centering
\addtolength{\tabcolsep}{2pt}
\renewcommand{\arraystretch}{1.5} 
\begin{tabular}{r|c|c}\hline \hline
\textbf{Basis} & \textbf{CFI} $\cI_\mathrm{NB}$ & \textbf{CFI} $\cI_\mathrm{BA}$\\ \hline 
Position & $0$ & $0$ \\
Momentum & $|\vn_\perp|^2 \cG $ & $|\vc_\perp|^2 \cG + \mathcal{O}(\vartheta^2)$ \\
OAM & $|\vn_\perp|^2 \cG - \cO\left( \vartheta^2 \right)$ &
$2 \cG - \cO\left( \vartheta^2 \right)$ \\\hline \hline
\end{tabular}
\caption{Classical Fisher information for different measurements of the electron state after probing a magnetic spin with and without backaction}
    \label{tab:CFI:NB_BA}
\addtolength{\tabcolsep}{-2pt}
\renewcommand{\arraystretch}{1}
\end{table} 

\begin{figure}
    \centering
    \includegraphics[width=\linewidth]{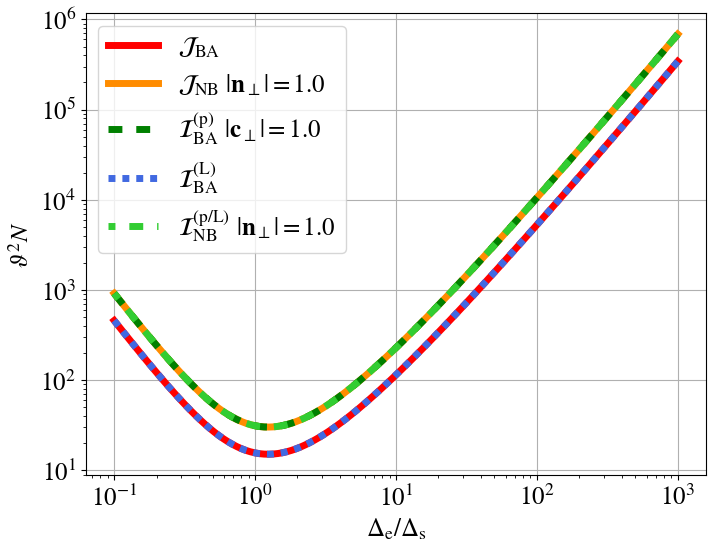}
    \caption{Required number of electrons to estimate a magnetic moment with $\mathrm{SNR}=3$, plotted as a function of the size ratio between the electron and the spin distributions. Both distributions are assumed Gaussian with standard deviations $\Delta_e$ and $\Delta_s$, respectively, yielding an optimal ratio of approximately $1.2$ (see text). We compare the numbers at a given weak interaction strength $\vartheta \ll 1$ for diffraction mode imaging (p) and binary OAM measurements (L) to the quantum limit set by the QFI, both with and without backaction. OAM measurements saturate the respective QFI bounds, whereas diffraction mode imaging saturates only the no-backaction QFI bound. A single electron spin spread over $\Delta_{\s} = 50\,$pm sourcing $\mu = \mu_{\rm B}$ amounts to $\vartheta \approx 5\times10^{-5}$ and $N> 10^{10}$ required probe electrons, a nuclear spin with $\Delta_{\s} = 1\,$pm amounts to $\vartheta \approx 2 \times 10^{-6} $ and $N>10^{13}$, and a magnetic sample of $10^5\mu_{\rm B}$ spread over $100\,$nm corresponds to $\vartheta \approx 3 \times 10^{-3}$ and $N>10^7$. 
  }
    \label{fig:Ne_sensing_Fisher}
\end{figure}

Table \ref{tab:CFI:NB_BA} summarizes the CFIs with and without backaction for the three types of measurement we considered.
Via the Cram\'{e}r-Rao bound \eqref{eq:CramerRaoBound}, the CFIs determine the highest signal-to-noise ratios one can achieve when estimating a parameter value $\vartheta$ with $N\gg 1$ single-electron shots. Conversely, a desired SNR at a given $\vartheta$-magnitude  sets the mimimum required number of probe electrons as
\begin{equation}\label{eq:N_SNR}
    N \geq \frac{\SNR^2}{\vartheta^2 \cI(\vartheta)} \geq \frac{\SNR^2}{\vartheta^2 \cJ(\vartheta)}.
\end{equation}
In Fig.~\ref{fig:Ne_sensing_Fisher}, we plot these CFI and QFI bounds at $\SNR=3$ as a function of the ratio $\chi = \Delta_{\e}/\Delta{\s}$ of probe focus to spin size for our Gaussian model. According to the QFI bounds at optimal focus, more than $\sim 10/\vartheta^2$ electrons are needed to reach the desired SNR.
The graph also illustrates the attainability of the quantum sensing limit: in the backaction case, an OAM measurement requires only half the number of electrons compared to the diffraction mode or OAM measurements in the no-backaction regime.

\section{Magnetic moment discrimination}\label{Sub:Basics:StateDiscrimination}

Our second, more basic, metrological task addresses the experimental challenge to detect single quantum spins: determining whether a magnetic sample was present or not. We are thus tasked with discriminating two distinct hypotheses: ($A$) the electron evolves freely without any interaction, and ($B$) the electron interacts with a magnetic dipole. These correspond to two distinct electron states in our scattering model, $\orho_A = \orho_{\e} (0)$ and $\orho_B = \orho_{\e} (\vartheta)$, which are then subjected to a POVM-measurement to produce random outcomes $\xi$ with the respective likelihoods, $P^{(A)}(\xi) = P(\xi|0)$ and $P^{(B)}(\xi) = P(\xi|\vartheta)$.

Given the random outcomes $\vxi$ from $N\gg 1$ independent shots, the goal is to decide which of the two likelihoods the outcomes were sampled from. For each single shot with outcome $\xi$, if we take $A$ and $B$ as equally likely a priori, the optimal strategy is to attribute $A$ whenever $P^{(A)}(\xi)\geq P^{(B)}(\xi)$ or $B$ otherwise. The success probability of attributing the correct hypothesis to the outcome, $P_{\rm suc} = [1 + D(P^{(A)}, P^{(B)})]/2$, is determined by the  `classical' trace distance between the two likelihoods \cite{nielsen2010quantum},
\begin{equation}
    \label{eq:Basics:ClTrDist}
    D(P^{(A)}, P^{(B)}) = \frac{1}{2} \int \diff \xi \left\vert P^{(A)}(\xi) - P^{(B)}(\xi) \right\vert.
\end{equation}
The greater it is, the more distinguishable the two likelihoods are;  perfect single-shot distinguishability, $D=1$, required them to be disjoint. Conversely, in the weak-sample regime, the two distributions barely differ and $D\ll 1$. Expanding $P^{(B)}$ to first order in $\vartheta$ and applying the Cauchy-Schwarz inequality provides a bound that links the discrimination task to the previous estimation task ,
\begin{equation}\label{eq:ClTrDist_CFIbound}
  D(P^{(A)}, P^{(B)}) \leq \frac{\vartheta}{2} \sqrt{\cI(0)} + \Order (\vartheta^2).  
\end{equation}
This bound is however not tight, as we will see in our case studies.

As before, we obtain the quantum limit for the success probability by optimizing the trace distance over all possible POVMs that could distinguish $\orho_A$ and $\orho_B$. The so defined quantum trace distance takes a simple form,
\begin{equation}\label{eq:Basics:QTrDist}
    D^\mathrm{Q}(\orho_A, \orho_B) = \frac{1}{2} \tr \left\{ \left| \orho_A - \orho_B \right| \right\},
\end{equation}
and the associated optimal quantum strategy is given by a binary projective POVM, $\cE^Q = \{ \oF, \mathds{1} - \oF \}$, where $\oF$ is the projector onto the positive eigenspace of $\orho_A - \orho_B$~\cite{nielsen2010quantum}. An upper bound can be obtained from \eqref{eq:ClTrDist_CFIbound} since the QFI bounds the CFI for any measurement, 
\begin{equation}\label{eq:QTrDist_QFIbound}
  D^\mathrm{Q} (\orho_A, \orho_B) \leq \frac{\vartheta}{2} \sqrt{\cJ(0)} + \Order (\vartheta^2).  
\end{equation}

The single-shot $P_{\rm suc}$ at hand, the number of correct attributions over $N\gg 1$ shots is an approximately Gaussian random variable with mean value $N_{\rm suc} = NP_{\rm suc} $ and standard deviation $\Delta = \sqrt{NP_{\rm suc}(1-P_{\rm suc})}$. In order to achieve a desired confidence level $\CL$ (e.g.~$87\%$) on average, the mean value must exceed $N/2$ by the respective confidence half-interval, $N_{\rm suc}-N/2 \geq \Delta \zeta$ with $\zeta = \sqrt{2}\erf^{-1}(\CL)$ (e.g.~$\zeta = 1.5$). This requires an  average number of shots of at least
\begin{equation}\label{eq:N_CL}
    N \geq 
    \left[\frac{1}{D^2} -1\right]\zeta^2 \geq \left[\frac{1}{(D^Q)^2} -1\right]\zeta^2 .
\end{equation}
Inserting the above CFI and QFI bounds on the trace distance, \eqref{eq:ClTrDist_CFIbound} and \eqref{eq:QTrDist_QFIbound}, and assuming they are much smaller than unity, the required shots for discrimination in \eqref{eq:N_CL} match those for estimation in \eqref{eq:N_SNR} with $\SNR = 2 \zeta$.

\subsection{Discrimination without backaction}\label{Sec:Met:DiscNB}

We first consider the NB case in which the magnetic moment is a fixed vector quantity of magnitude $\mu$ and in-plane component $\vn_{\perp}$. Upon scattering, the probe electron remains in a pure state, $|\psi(0)\ra \to |\psi(\vartheta)\ra$, and with our assumption that the incident wavefunction is radially symmetric, the quantum trace distance can be readily calculated. We find,
\begin{align}\label{eq:DQ_NB}
D_{\rm NB}^\mathrm{Q} \left( \ket{\psi(0)}, \ket{\psi(\vartheta)} \right) 
    &=\sqrt{1-\la \psi(0)| J_0 \left[ \vartheta |\vn_\perp|g(\opr) \right] |\psi(0)\ra^2},\nonumber \\
    &= \frac{\vartheta}{2}|\vn_\perp| \sqrt{\cG} + \cO\left( \vartheta^3\right),
\end{align}
which saturates the QFI bound \eqref{eq:QTrDist_QFIbound} in the weak-sample limit. Hence, the conditions for optimal sensitivity in Sec.~\ref{Sec:Sensing_NB} carry over; in particular, the electron should be focused down to about the size of the magnetic sample. However, attaining the quantum-limited discrimination probability would require a projective measurement onto the positive eigenspace of $\ket{\psi(\vartheta)} \bra{\psi(\vartheta)} - \ket{\psi(0)} \bra{\psi(0)}$, inaccessible in practice. See App.~\ref{App:QTraceDerivation} for details.

We find that neither the OAM measurement nor conventional TEM imaging can saturate the quantum bound \eqref{eq:DQ_NB}, in contrast to the estimation task in Sec.~\ref{Sec:Sensing_NB}. In fact, the binary OAM measurement,  optimal for the \textit{estimation} of small $\vartheta$ over many shots, turns out to be a comparably poor strategy for \textit{discriminating} $\vartheta$ from zero in each single shot. The classical trace distance \eqref{eq:Basics:ClTrDist} for the binary measurement can be expressed in terms of the zero-OAM likelihood \eqref{eq:zeroOAM_P},
\begin{equation}
    D^{\rm (L)}_{\rm NB} = 1 - \la \psi(0)| J_0^2 [ \vartheta |\vn_{\perp}| g(\opr)] |\psi(0) \ra = \left| \frac{\vartheta \vn_{\perp}}{2}\right|^2 \cG + \Order (\vartheta^4).
\end{equation}
To leading order in $\vartheta$, it is given by the square of the quantum trace distance \eqref{eq:DQ_NB} and therefore far from optimal. This holds true for arbitrary $\vartheta$-values, because $P_0(\vartheta)\geq 1- (D_{\rm NB}^{\rm Q})^2$ and thus $ D_{\rm NB}^{\rm (L)} \leq (D_{\rm NB}^\mathrm{Q})^2$ in general.

For conventional TEM imaging, let us consider a variable setup ranging from image mode to diffraction mode, as the latter is no longer optimal for discrimination. We assume that the paraxial electron beam is imaged through an ideal thin lens of focal length $f$ at distance $z$ onto a perfectly resolving detector screen at distance $2z$ from the sample. This corresponds to a measurement of the transverse electron state in the basis of
\begin{equation}\label{eq:xiz_basis}
    |\vxi_z\ra = \exp{\im \frac{z\lambda_0 \ovp_{\perp}^2}{4\pi \hbar^2}}  \exp{ \im\frac{ \pi \ovx_{\perp}^2}{f\lambda_0}} \exp{\im \frac{z\lambda_0\ovp_{\perp}^2}{4\pi \hbar^2}} |\vx_{\perp}\ra, 
\end{equation}
with $\lambda_0 = 2\pi\hbar/p_0$ the electron wavelength. We allow the distance to vary from $z=0$ to $f$, which correspond to measuring $\vxi_0=\vx_{\perp}$ (image mode) and $\vxi_f = f \vp_{\perp}/p_0$ (diffraction mode), respectively. For small $\vartheta$, the associated likelihood functions, $P(\vxi_z|\vartheta) = |\la \vxi_z|\psi(\vartheta)\ra|^2$, and trace distances can still be expressed in terms of only a few integrals. Introducing polar detector coordinates, $\vxi_z = (r,\varphi)$ and expanding the scattered amplitude to first order in $\vartheta$,
\begin{equation}
    \la \vxi_z|\psi(\vartheta)\ra = \la \vxi_z|\psi(0)\ra + \vartheta |\vn_{\perp}|\Xi_z(r) \sin\varphi  + \Order(\vartheta^2),
\end{equation}
we obtain the leading-order trace distance:
\begin{equation}
        D_{\rm NB}^{(z)} = 4 \vartheta |\vn_\perp| \int_0^\infty \!\!\! r\diff r \, \left| \Re \left\{ \langle \psi(0) | \vxi_z \rangle \, \Xi_z(r) \right\} \right| + \mathcal{O}(\vartheta^2).
\end{equation}
In the Gaussian model, the term $\Xi_z (r)$ reduces to an analytic, albeit cumbersome, expression. Detailed calculations are presented in App.~\ref{App:NB:First_order_defocus}. In image mode ($z=0$), the trace distance vanishes for arbitrary incident states and $\vartheta$, because position measurements are insensitive to position-dependent phases caused by the sample. In diffraction mode ($z=f$), we have that $D_{\rm NB}^{\rm (p)} = \vartheta |\vn_{\perp}|\cD + \Order(\vartheta^2)$, with
\begin{equation}\label{eq:D}
    \mathcal{D}=\sqrt{\frac{2}{\pi}}  \frac{1}{\chi} \left( \sqrt{\frac{1 + 2\chi^2}{1 + \chi^2}} - 1 \right).
\end{equation}
Compared to the quantum trace distance \eqref{eq:DQ_NB}, this is reduced by the factor $2\cD/\sqrt{\cG} \leq \sqrt{2/\pi} \approx 0.8$. The factor approaches its maximum value $0.8$ in the limit of a tightly focused beam, $\chi \ll 1$, whereas it vanishes logarithmically for $\chi \gg 1$. At the optimal focus ($\chi \approx 1.2$), we get a moderate  reduction of about $0.6$.

\begin{figure*}
    \centering
    \includegraphics[width=\linewidth]{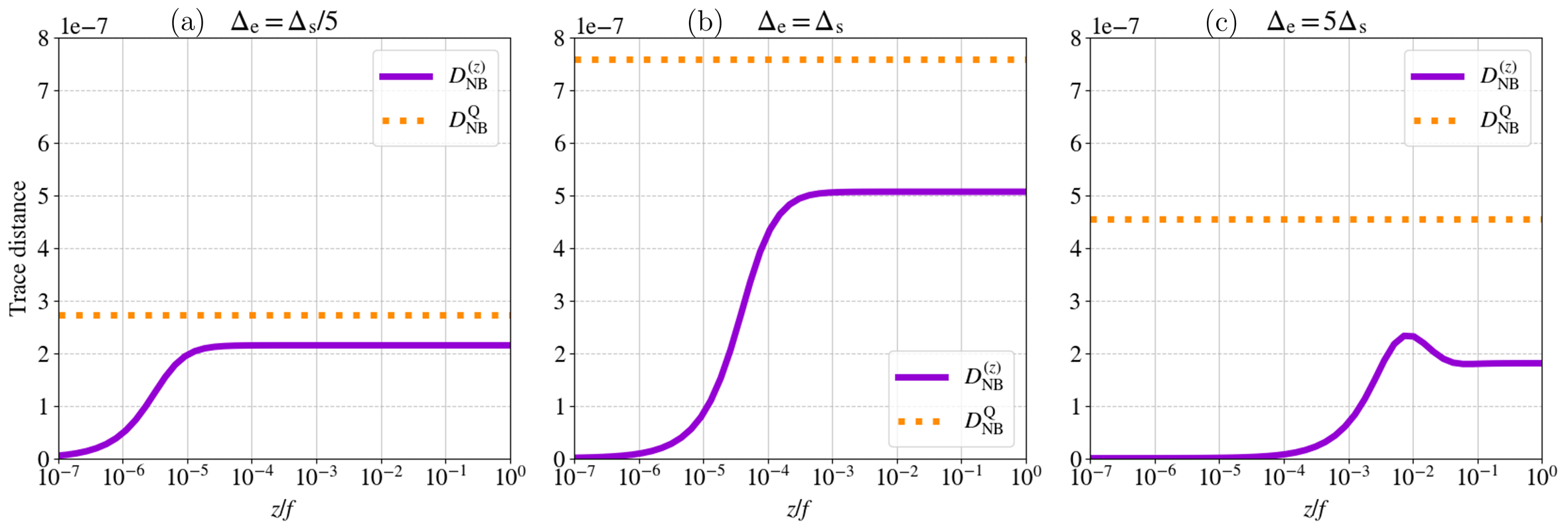}
    \caption{Classical trace distance for magnetic moment discrimination without backaction in TEM imaging at varying distance parameter $z$, from image mode ($z=0$) to diffraction mode ($z=f$). We assume a magnetic moment $\mu = 100\mu_{\rm B}$ spread over a Gaussian of size $\Delta_\s = 1.0\,$nm and $200\,$keV electrons of wavelength $\lambda_0 = 2.5\,$pm with three Gaussian focus sizes: (a) $\Delta_\e = \Delta_\s/5$, (b) $\Delta_\e = \Delta_\s$, (c) $\Delta_\e = 5 \Delta_\s$. The settings result in a phase parameter $\vartheta=2.8\cdot 10^{-4}$ and quantum trace distances marked by the orange dotted lines. The focal length of the imaging lens is taken as $f=2.0\,$mm. 
    }
    \label{fig:Disc:DefocusPlanes}
\end{figure*}

We showcase the performance of magnetic moment discrimination with TEM imaging in Fig.~\ref{fig:Disc:DefocusPlanes}, exemplarily for a magnetic sample with $\mu = 100\mu_{\rm B}$ and typical beam parameters in the Gaussian model without backaction. We plot the trace distance as a function of the distance parameter $z$ for (a) a tight beam focus, (b) a nearly optimal focus, and (c) a wide focus. In all three cases, we observe that the diffraction-mode value determined by \eqref{eq:D} is quickly approached already for defocus plane imaging at $z\ll f$; in (c), one can even exceed the diffraction mode at $z\sim 10^{-2} f$, albeit at a lower overall performance compared to the quantum limit (red dotted lines). We attribute this enhancement to interference effects between the wide unscattered beam and the scattered component.

\begin{figure}
    \centering
    \includegraphics[width=1\linewidth]{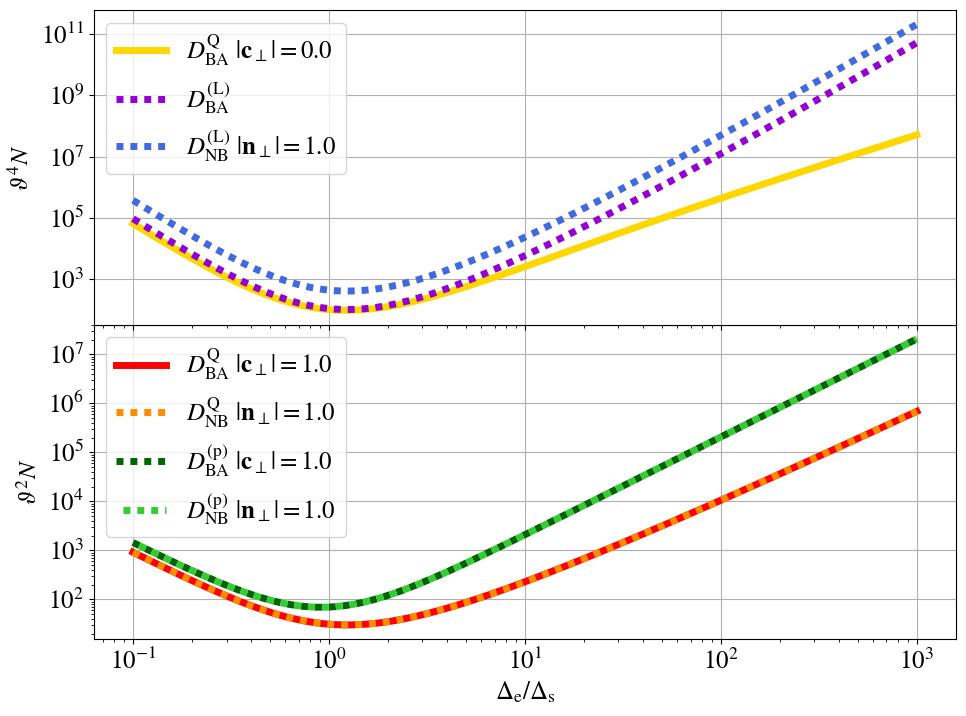}
    \caption{Required number of shots to detect a magnetic dipole in the electron path with $87\%$ confidence, set by the trace distance, as a function of the ratio between the probe size $\Delta_\e$ and the spin size $\Delta_\s$. We compare diffraction mode imaging and OAM measurements with and without backaction to the quantum bounds for a spin oriented longitudinally ($\vc_{\perp}=0$) and transversally to the beam; the two lowest curves are equal and resemble the no-backaction QFI bound in Fig.~\ref{fig:Ne_sensing_Fisher} at $\SNR \approx 3$. In all cases, the minimum is reached when the probe size approximately matches the spin size. Notice the different scaling of the number with $\vartheta^{-4}$ and $\vartheta^{-2}$ in the upper and lower panel, implying that many more shots are needed in the upper panel at the same $\vartheta \ll 1$.  
    }
\label{fig:Nshots_discrimination}
\end{figure}

\begin{table}
\addtolength{\tabcolsep}{2pt}
\renewcommand{\arraystretch}{1.5} 
\centering
\begin{tabular}{r|c|c}\hline \hline
\textbf{Basis} & Trace distance $D_\mathrm{NB}$ & Trace distance $D_\mathrm{BA}$ \\ \hline 
Position & $0$ & $0$\\
Momentum & ${\vartheta}|\vn_\perp| \mathcal{D} + \mathcal{O}(\vartheta^2)$ & ${\vartheta}|\vc_\perp| \mathcal{D} + \mathcal{O}(\vartheta^2)$ \\
OAM & $\frac{1}{4} \vartheta^2|\vn_\perp|^2 \cG + \cO\left( \vartheta^4 \right)$ & $\frac{1}{2} \vartheta^2 \cG + \cO\left( \vartheta^4 \right)$
\\\hline \hline
\end{tabular}
\caption{Classical trace distance between outcome likelihoods when discriminating a magnetic spin by different measurements on a probe electron with and without backaction.}
    \label{tab:TraceDist:NB_BA}
\addtolength{\tabcolsep}{-2pt}
\renewcommand{\arraystretch}{1} 
\end{table}

\subsection{Discrimination with backaction}\label{Sec:Met:DiscBA}

The discrimination task is especially relevant when it comes to sensing the magnetic moment of single quantum spins. The backaction caused by the electron  then plays an important role; in particular, the scattered electron state \eqref{eq:finalstate_BA} is now mixed, which complicates the evaluation of the trace distance. Nevertheless, in the weak-sample limit, the quantum trace distance behaves similarly as in the no-backaction case,
\begin{equation}\label{eq:DiscBA:QtrDist}
D^{\rm Q}_{\rm BA}=\begin{cases}
    \frac{1}{2}\vartheta|\vc_\perp|\sqrt{\cG} +\cO(\vartheta^2), & \text{if } |\vc_\perp|>\frac{1}{\sqrt{8}}\vartheta,\\
    \frac{1}{4}\vartheta^2 \left[\cG + \cF \right]
    & \text{if } |\vc_\perp|=0,
\end{cases}
\end{equation}
as follows from a lengthy calculation in App.~\ref{App:QTrDist}. In contrast to the NB case \eqref{eq:DQ_NB}, the quantum trace distance now does not vanish when $|\vc_\perp| = 0$, but instead is of second order in $\vartheta$. Here, $\cF = 2\sqrt{\la\psi(0)|g^4(\opr)|\psi(0)\ra} \geq \cG$ 
in general.  
For the Gaussian model, $\cF$ can be expressed explicitly in terms of $\chi$, as shown in Eq.~\eqref{eq:G4}. The maximum of $\cF+\cG \approx 0.61$ is also assumed at $\chi\approx 1.2$.

The classical trace distance for the binary OAM measurement is again determined by the zero-OAM likelihood \eqref{eq:zeroOAM_PBA}. As in the NB-case, we find that it is of second $\vartheta$-order in the weak-sample limit,
\begin{equation}\label{eq:D_OAM_BA}
    D_\mathrm{BA}^{\rm (L)} = \la \psi(0)|\sin^2 \left[ \vartheta g(\opr) \right] |\psi(0)\ra = \frac{\vartheta^2}{2} \cG + \cO \left( \vartheta^4 \right).
\end{equation}
The OAM measurement is largely unsuitable to discriminate weak magnetic moments, unless for $\vc_{\perp}=0$, when \eqref{eq:D_OAM_BA} almost saturates the quantum trace distance, $D_\mathrm{BA}^{\rm (L)}/D_\mathrm{BA}^{\rm Q} \approx 0.97$  at $\chi\approx 1.2$ in the Gaussian model.

For conventional TEM imaging, we restrict our view to diffraction mode, since we could already see in the NB-case that there is hardly any room for improvement with defocus plane imaging; and image mode is insensitive to $\vartheta$. Expanding the momentum likelihood, $P(\vp_{\perp}|\vartheta) = \la \vp_{\perp}| \orho_\e (\vartheta)| \vp_{\perp}\ra$, in $\vartheta$, we arrive at $D_\mathrm{BA}^{\rm (p)} = \vartheta |\vc_\perp| \mathcal{D} + \mathcal{O}( \vartheta^2 )$, as shown in Appendix~\ref{App:DiscBA:PosAndMom}.

In summary, the discrimination of weak magnetic moments can be equally successful with and without backaction. Given the same in-plane spin magnitude, $|\vn_{\perp}|=|\vc_{\perp}|$, the relevant trace distances match, with the exception of $|\vn_{\perp}|,|\vc_{\perp}| = 0$. Here, the quantum backaction effect retains a certain (second-order) sensitivity of the electron (and its OAM state) to $\vartheta$, which would be absent  without backaction. Table~\ref{tab:TraceDist:NB_BA} summarizes the trace distances for the three measurement types considered.

In Fig.~\ref{fig:Nshots_discrimination}, we plot the required number of shots $N$ to detect the presence of a weak magnetic moment at $\CL=87\,\%$ as a function of the ratio $\chi = \Delta_{\e}/\Delta{\s}$ in the Gaussian model. We compare the quantum bounds to the classical ones for OAM and momentum measurements. 
Given $\vartheta \ll 1$, the bounds on $N$ in Eq.~\eqref{eq:N_CL} can be approximated as $N\geq (\zeta/D)^2 \geq (\zeta/D^{\rm Q})^2$,
the bounds set by $D^{\rm Q}$ scale as $\vartheta^{-2}$ and are inversely proportional to the in-plane component 
($|\vn_{\perp}|$, or $|\vc_{\perp}|$ when $|\vc_{\perp}| > \vartheta/\sqrt{8}$). With or without backaction, diffraction-mode imaging with well-focused beams ($\Delta_\e \lesssim \Delta_\s$) operates close to the quantum limit. At $\vc_\perp = 0$, detection is only possible at all in the case with backaction, but the quantum bound on $N$ scales as $\vartheta^{-4}$. The binary OAM measurement then achieves a similar performance for $\Delta_\e \lesssim 10\Delta_\s$.

\section{Discussion and Outlook}\label{Sec:Discussion}

We have presented a quantitative analysis of the metrological sensitivity of electron beams in a TEM to weak magnetic samples on the beam axis, when the goal is to discriminate the presence of a magnetic moment or to estimate its magnitude. We benchmarked the sensitivity achievable in conventional TEM imaging against fundamental quantum limits to sensitivity as well as a proposed measurement that detects the electron's orbital angular momentum. Analytic results could be obtained when the electron beam profile and the spatial distribution of the magnetic moment are both Gaussians centered on axis.

In particular, our analysis compares the standard scenario of a magnetic moment sourced by many spins (electron spins or nuclear spins), and thus unaffected by the probe electron, to the ultimate quantum regime in which the magnetic moment is sourced by a single quantum spin and backaction of the electron onto that spin state is significant. We found, somewhat counter-intuitively, that the quantum backaction effect may \textit{improve} sensitivity.
Without backaction, electrons are only sensitive to a transverse magnetic moment. In this case, we proved that diffraction mode imaging is optimal for estimation, while it is nearly optimal (factor $\sim 2$) for discrimination with a beam focused on the scale of the spin system.
With backaction, the sensitivity of diffraction mode imaging remains the same, but it no longer saturates the quantum bound. Instead, a measurement of the electron's OAM is optimal for estimation, improving the SNR by $\sqrt{2}$ or more. Moreover, the OAM does not depend on the spin orientation. Regarding discrimination, the OAM measurement is near-optimal only for spins without a transverse component and beam focused on the scale of the spin system. OAM detection schemes have recently been demonstrated for TEM \cite{tavabi2021OAMsorter}.

\begin{figure}
        \centering
        \includegraphics[width=\linewidth]{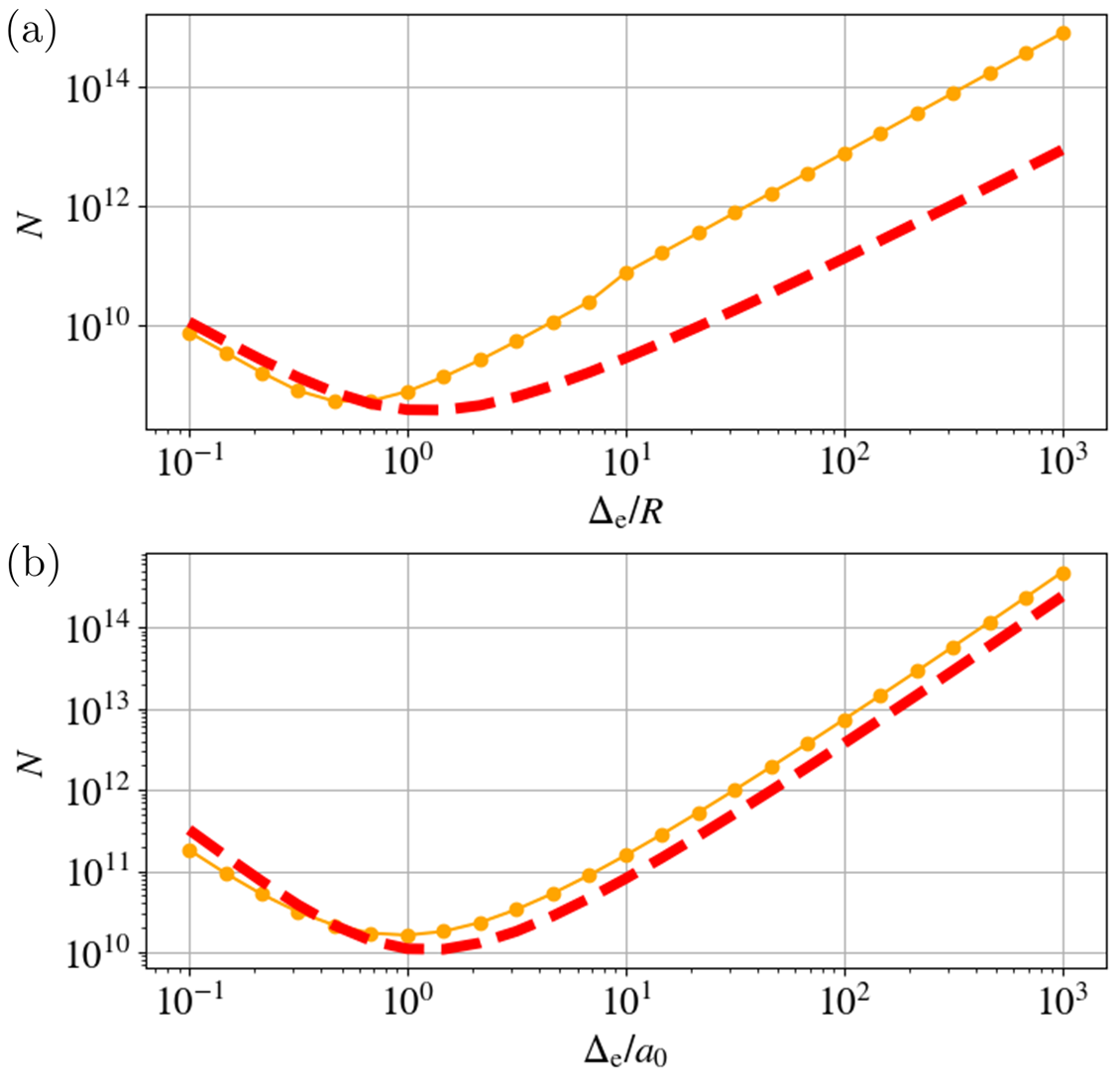}
        \caption{Quantum bounds on the required number of electrons to estimate realistic magnetic moments at $\SNR=3$, shown as a function of the probe size $\Delta_\e$. We consider (a) a superparamagnetic nanoparticle modeled as a solid sphere of radius $R=1$~nm with $\mu = 100\,\mu_{\rm B}$ and (b) a single electron spin ($\mu = \mu_{\rm B}$) delocalized over the $1s$ orbital of a hydrogen atom with Bohr radius $a_0 \approx 52$~pm, including backaction. In each case, we compare the exact evaluation of the QFI using said spin densities (orange) to the Gaussian model with $\Delta_\s = R$ and $a_0$ (red). The required electron number for the discrimination task coincides with those shown here for a CL=$87\%$. In the backaction regime, sensing requires a factor of two fewer electrons than discrimination.}
        \label{fig:realistic_NeNshots}
\end{figure}

An important question is to what extent the results of our Gaussian model still apply to magnetic moments sourced by more realistic spin distributions. To this end, we consider two benchmark examples in Fig.~\ref{fig:realistic_NeNshots}: (a) a fairly large magnetic moment of $\mu=100\mu_\mathrm{B}$ distributed homogeneously inside a superparamagnetic nanosphere~\cite{Bedanta_2009,Morup_2010, Gossuin_2009} of $R=1\,$nm radius, and (b) a single electron spin delocalized over the 1s-orbital of a hydrogen atom with Bohr radius $a_0 \approx 52\,$pm. The respective phase parameters \eqref{eq:phase} amount to $\vartheta \sim 10^{-4}$ and $10^{-5}$. The radial interaction profiles \eqref{eq:g} are detailed in App.~\ref{App:Real}. 
We treat the first example without backaction, assuming that the femtosecond-short transit of the probe electron cannot resolve the dense level structure or manipulate the quantum state of the magnetic spins sourcing $\mu$.

Figure \ref{fig:realistic_NeNshots} depicts the resulting quantum-limited number of single-electron shots required for magnetic moment estimation at $\SNR=3$ as a function of the Gaussian electron beam focus. The results are compared to those assuming Gaussian spin distributions of the same radii (red dashed). We observe that the Gaussian model closely matches the results for the hydrogen spin in (b) throughout, and it predicts the minimal shot number well in both cases. However, the position of the minimum and the behaviour at large beam foci deviates significantly in the nanosphere case (a). 

Example (b) suggests that one needs about $10^{10}$ probe electrons to detect a single atomic electron spin, assuming the probe beam is centered on the atom and focused to the size of the atomic orbital. For a single nuclear spin spread over, say, $\Delta_\s = 1\,$pm and an extreme electron beam focus of $\Delta_\e = 10\,$pm, one would need about $10^{14}$ shots; see App.~\ref{App:Real}.
Operating a scanning transmission electron microscope at a high beam current of about 
\SI{1.5}{\nano\ampere} and sub-\AA ngstr{\"o}m spatial resolution would result in approximately $10^{10}$ electrons per second, which suffices to sense single electron spins within a few seconds of averaging. Inherent magnetic fields of up to $\SI{2}{\tesla}$ could serve as a bias field to achieve $30\%$ spin polarization at a temperature of $\SI{4}{\kelvin}$.

In contrast, nuclear spins, which are much more abundant, must be cooled to lower temperatures for a similar level of spin polarization: approximately \SI{7}{\milli\kelvin} 
for 30\% nuclear spin polarization at \SI{2}{\tesla}, alternatively hyperpolarisation techniques could be applied \cite{Capozzi2015}. Under such conditions, only $10^{10}$ electrons would be needed to sense the nuclear spin state of individual atomic column consisting of hundreds of aligned atoms along the beam direction in typical 100-200 nm thick samples. 
Importantly, nuclear spins exhibit extremely long coherence times, ranging from minutes to even hours~\cite{Callaghan1993Principles}, and the spin precession frequency of approximately $42$ MHz$/$T is slow enough to directly time-resolve the nuclear spin dynamics using fast direct electron cameras \cite{Llopart_2022}.
A recent proposal for electron number squeezing~\cite{evenhaim2025spinsqueezingelectronmicroscopy} suggests that the required number of electrons could be further reduced by approaching a Heisenberg scaling of the SNR with the electron number, paving the way to the sensing of single nuclear spins.

The scattering metrology framework could be extended beyond radially symmetric magnetic samples, to aloof electron beams, and to multiparameter sensing of additional parameters such as spin orientation and locations of magnetic domains. One could also consider more general input states such as spin-polarized or OAM electron beams, tailored for maximum sensitivity to specific samples.
Moreover, the framework could also accommodate a greater variety of interactions between free electrons and quantum systems, such as electrostatic or spin-spin interactions.

\acknowledgments

M.G.~acknowledges support from the House of Young Talents of the University of Siegen. We thank Benjamin Yadin, Anton{\'i}n Jar{\v o}s, Michael Seifner, and Stefan Rotter for helpful discussions. PH thanks the Austrian Science Fund (FWF): P36041, P35953, Y1121 and the FFG project AQUTEM.

%



\hypertarget{sec:appendix}
\appendix

\renewcommand{\thesubsubsection}{\Alph{section}.\Roman{subsection}.\arabic{subsubsection}}
\renewcommand{\thesubsection}{\Alph{section}.\Roman{subsection}}
\renewcommand{\thesection}{\Alph{section}}
\setcounter{equation}{0}
\numberwithin{equation}{section}
\setcounter{figure}{0}
\numberwithin{figure}{section}
\renewcommand{\theequation}{\Alph{section}.\arabic{equation}}
\renewcommand{\thefigure}{\Alph{section}.\arabic{figure}}


\newpage
\onecolumngrid

\newpage
\section{Derivation of the electron--spin Hamiltonian}\label{App:HamDerivation}

Here we derive the regularized electron-sample Hamiltonian \eqref{eq:Met:RegHam}. We start with the vector potential of a magnetic dipole $\vmu$ regularized over a radially symmetric spin density, $P_\mathrm{s}(\vu) = P_\mathrm{s}(|\vu|)$:
\begin{equation}\label{eq:app:regularized_vectorField}
    \vA_\mathrm{s}(\vx) = \int \diff^3 \vu \, P_\mathrm{s}(|\vx-\vu|) \vA_0(\vu) 
    = \frac{\mu_0}{4 \pi} \vmu \times \int \diff^3 \vu \frac{P_\mathrm{s}(|\vx - \vu|)}{|\vu|^3}\vu = \frac{\mu_0}{4 \pi} \vmu \times \frac{2\pi\vx}{|\vx|} \int_0^\infty \!\!\! \diff u \int_{-1}^1 \!\!\diff \tau \, \tau P_{\s} \left( \sqrt{u^2 + |\vx|^2 - 2u|\vx|\tau} \right) .
\end{equation}
In the final step, we represent $\vu = \vu(u,\theta,\varphi)$ in spherical coordinates with respect to $\vx$ and abbreviate $\tau = \cos\theta$, such that $\vu\cdot\vx = u \tau |\vx|$ and $\int_0^{2\pi}\diff\varphi \, \vu = 2\pi u \tau \vx/|\vx|$. One can now see that the regularized vector potential is simply the unregularized one multiplied by the function \eqref{eq:F}, $\vA_{\s} (\vx) = F(|\vx|) \vA_0 (\vx)$. We remark that, if the magnetic moment arises from a quantum particle with spin, we assume that the spatial wavefunction of that particle remains unaffected by the interaction with the probe electron and the spin density $P_{\s}$ is thus static.

For a Gaussian density, $P_{\s} (u) = \exp(-u^2/2\Delta_{\s}^2)/(2\pi\Delta_{\s}^2)^{3/2}$, the regularizer \eqref{eq:F} simplifies to
\begin{equation}
\label{eq:App:radialF_function}
F(x) = \frac{2\pi x^2}{(2\pi \Delta_\s^2)^{3/2}} \int_0^\infty \diff u \int_{-1}^1 \diff \tau \, \tau \,  \exp\left(-\frac{u^2 + x^2 - 2ux\tau}{2\Delta_\s^2} \right) = \mathrm{erf}\left( \frac{x}{\sqrt{2} \Delta_\s} \right) - \sqrt{\frac{2}{\pi}} \frac{x}{\Delta_\s} \exp\left( - \frac{x^2}{2\Delta_\s^2} \right).
\end{equation}
To arrive at the right hand side, we first integrate over $\tau$ and then use the identity:
\begin{align}
     \int_0^\infty \frac{\diff u}{u^2} \exp(-b u^2)\left[ \sinh(uc) -u c \cosh(uc) \right] &= c - \sqrt{\pi b} \exp\left( \frac{c^2}{4b} \right) \mathrm{erf} \left( \frac{c}{2\sqrt{b}} \right).
\end{align}
Near the origin, $F(x)$ admits a series expansion, $F(x) = \sqrt{2/9\pi} (x/\Delta_\s)^3 + \mathcal{O} [(x/\Delta_\s)^5 ]$, 
whereas for $x \gg \Delta_\s$ it approaches unity. Hence, the regularization introduces a smooth cutoff at short distances without altering the long-range behavior of the magnetic vector potential.

Another useful property of the vector potential is that its operator form $\vA_{\s} (\ovx)$ commutes with the momentum operator $\ovp$. To prove this, we first notice that the vector potential is of the form $\vA_{\s} (\vx) = \vmu \times \nabla M(|\vx|)$, with some scalar function $M$ that obeys $M'(x) = \mu_0 F(x)/4\pi x^2$. From this follows,
\begin{equation}\label{eq:App:conmutator_pA}
    \vA_{\s}(\ovx) \cdot \ovp - \ovp \cdot \vA_{\s}(\ovx) = 
    \sum_i \left[A_{\s,i}(\ovx), \op_i \right]
    = \sum_{i,j,k} \epsilon_{ijk} \mu_j \left[ \partial_k M(|\ovx|), \op_i \right] = \im \hbar  \sum_{i,j,k} \epsilon_{ijk} \mu_j \partial_i\partial_k M(|\ovx|) = 0,
\end{equation}
which implies $\{ \vA_{\s}(\ovx), \ovp \} = 2 \vA_{\s}(\ovx) \cdot \ovp$.
This anticommutator appears when we expand the minimal coupling Hamiltonian \eqref{eq:Met:MinCoup} to lowest order around the free electron energy. The magnetic bias field also contributes a term to the vector potential, $\vA(\ovx) = \vA_{\s} (\ovx) + \vB_0 \times \ovx/2$, which leads to a deflection of the electron beam of similar order as that induced by the spin field. However, this effect appears only as a background on top of the spin-induced signal and can be safely ignored, since its deflection is either compensated by the electron optics (lenses) or removed in practice by image subtraction, where constant contributions not related to spin dynamics are eliminated. Omitting the negligible $\vA^2$-contribution, we arrive at
\begin{equation}\label{eq:App:RelHAM_minimalCoup_approx}
        \left[(mc^2)^2+c^2\left( \ovp + e\vA_\s(\ovx) \right)^2\right]^{1/2} 
        \approx mc^2 \gamma(|\ovp|)
        +\frac{ e \vA_\s(\ovx)\cdot \ovp}{m\gamma(|\ovp|)} ,
\end{equation}
where $\gamma(|\ovp|)=\sqrt{1+|\ovp|^2/(mc)^2}$ is the Lorentz factor. 
As is standard in electron microscopy, the leading relativistic correction simply increases the effective electron mass by $\gamma$~\cite{1961_fujiwara}. Note that, on the right hand side, the correct operator ordering must be specified, for which we refer the reader to Ref.~\cite{simulationSRS_TEM}. To consistent leading order, we can take the symmetric form,  
\begin{equation}
    \oV=\frac{ e \,\ovA_\s(\ovx)\cdot \ovp}{m\gamma(|\ovp|)}
    =\frac{e}{2m}\left(\ovA_\s(\ovx)\cdot \frac{\ovp}{\gamma(|\ovp|)}+\frac{\ovp}{\gamma(|\ovp|)}\cdot \ovA_\s(\ovx)\right).
\end{equation}
The remaining Lorentz factor will eventually cancel out in the paraxial scattering operator, as we will see in App.~\ref{App:Met:ScattSVEA}.

\section{Derivation of the paraxial scattering operator}
\label{App:Met:ScattSVEA}

\subsection{Scattering Operator Without Magnetic Dipole Backaction}\label{App:Met:ScattNoBa}

We begin with the Hamiltonian $\oH_\mathrm{NB}$, which is defined from Eq.~\eqref{eq:Met:RegHam} with $\vmu=\mu\vn$, as follows
\begin{equation}\label{eq:Met:RegHam_NB}
        \oH_\mathrm{NB} = \sqrt{(mc^2)^2+c^2\left| \ovp\right|^2} + \oV_{\rm NB} ,\qquad \text{with} \quad 
        \oV_{\rm NB} := \frac{\hbar r_\mathrm{e}}{2m} \left( \frac{\mu}{\mu_\mathrm{B}} \right) \vn \cdot \left( \frac{F(|\ovx|)}{|\ovx|^3} \frac{\ovx\times \ovp}{\gamma(|\ovp|)} \right),
\end{equation}
which describes the interaction of a free electron with a magnetic field.  
To simplify the dynamics, we adopt the slowly varying envelope approximation (SVEA)~\cite{BONIFACIO1993185} using the ansatz
\begin{equation}\label{eq:App:boosted_wavefucntion}
    \ket{\psi(t)} = \exp \left( \im \frac{\vp_0 \cdot \ovx - E_0 t}{\hbar} \right) \ket{u(t)}, 
    \qquad E_0=\sqrt{(mc^2)^2+c^2\vp_0^2},
\end{equation}
which transforms the time-dependent Schr{\"o}dinger equation into
\begin{align}
    \im \hbar \frac{\diff}{\diff t} \ket{u(t)} 
    &= \left[ \exp\left( -\im\frac{\vp_0 \cdot \ovx}{\hbar} \right) 
    \oH_\mathrm{NB} 
    \exp\left( \im\frac{\vp_0 \cdot \ovx}{\hbar} \right) - E_0 \right] \ket{u(t)} \\
    &= \Bigg[ \sqrt{(mc^2)^2+c^2(\ovp+\vp_0)^2} 
    + \frac{\hbar r_\mathrm{e}}{2m} \left( \frac{\mu}{\mu_\mathrm{B}} \right) 
    \vn \cdot \left(\frac{F(|\ovx|)}{|\ovx|^3} 
    \frac{\ovx\times (\ovp + \vp_0)}{\gamma(|\ovp+\vp_0|)} \right) 
    - E_0 \Bigg] \ket{u(t)} .
\end{align}
Assuming $|\ovp| \ll |\vp_0|$ and neglecting higher-order terms in $\ovp$, we expand as
\begin{subequations}
    \begin{align}
\sqrt{(mc^2)^2+c^2(\ovp+\vp_0)^2}
    &\approx \sqrt{(mc^2)^2+c^2\vp_0^2}
    +\frac{\vp_0\cdot \ovp}{\gamma(|\vp_0|)m}+\frac{\ovp^2}{2\gamma(|\vp_0|)m}, \\[6pt]
\frac{\ovp+\vp_0}{\gamma(|\ovp+\vp_0|)}
    &\approx \frac{\vp_0}{\gamma(|\vp_0|)}
    -\frac{\vp_0(\ovp\cdot \vp_0)}{\gamma^3(\vp_0)m^2c^2}
    +\frac{\ovp}{\gamma(|\vp_0|)},
    \end{align}
\end{subequations}
where, in the second equation, the combined contribution of the second and third terms scales as $|\ovp|/\gamma^3(|\vp_0|)$.  
Thus, to leading order, only the first term is retained.
We then obtain the effective dynamics
\begin{equation}\label{eq:App:timeDependentU_SVEA}
    \begin{split}
\im \hbar \frac{\diff}{\diff t} \ket{u(t)} \approx&
\Bigg[\frac{\vp_0}{m \gamma(|\vp_0|)} \cdot \ovp+\frac{\ovp^2}{2\gamma(|\vp_0|)m} 
+ \frac{\hbar r_\mathrm{e}}{2m} \left( \frac{\mu}{\mu_\mathrm{B}} \right) 
\vn \cdot \left( F(|\ovx|) \frac{\ovx}{|\ovx|^3} \times \frac{\vp_0}{\gamma(|\vp_0|)} \right) \Bigg] \ket{u(t)} \\
=& \left[\underbrace{v_0 \op_z +\frac{\ovp^2}{2\gamma(|\vp_0|)m}}_{\oH_{\rm NB}^{(0)}}
+ v_0 \frac{\hbar r_\mathrm{e}}{2} \left( \frac{\mu}{\mu_\mathrm{B}} \right)  
F(|\ovx|) \frac{n_x \oy - n_y \ox}{|\ovx|^3}  \right] \ket{u(t)},
    \end{split}
\end{equation}
where, in the second line, we set $\vp_0 = p_0 \ve_z$ and defined $v_0 = p_0 / \big(m\gamma(|\vp_0|)\big)$.  
The first two terms in Eq.~\eqref{eq:App:timeDependentU_SVEA} define $\oH_{\rm NB}^{(0)}$ and represent the free evolution term, while the third corresponds to the interaction Hamiltonian, which in the interaction-picture, becomes
\begin{equation}
\begin{split}
       \oV_{\rm NB}^{(\mathrm{I})}(t) =& \e^{\im \oH_{\rm NB}^{(0)} t/\hbar} \oV_{\rm NB} \e^{-\im \oH_{\rm NB}^{(0)} t/\hbar} 
   = \frac{\hbar r_\mathrm{e}}{2} \left( \frac{\mu}{\mu_\mathrm{B}} \right) v_0 \left( n_x \oy - n_y \ox \right) \frac{F\left( \left|\ovx + v_0t \ve_z+\frac{\ovp}{2\gamma(|\vp_0|)m}t\right|  \right)}{\left|\ovx + v_0t \ve_z+\frac{\ovp}{2\gamma(|\vp_0|)m}t\right|^3},\\
   \approx& \frac{\hbar r_\mathrm{e}}{2} \left( \frac{\mu}{\mu_\mathrm{B}} \right) v_0 \left( n_x \oy - n_y \ox \right) \frac{F\left( |\ovx + v_0t \ve_z|  \right)}{|\ovx + v_0t \ve_z|^3}.
\end{split}
\end{equation}
In the last step, we neglected the contribution from transverse momentum and the deviation in the longitudinal momentum from $p_0$, which is much smaller than $p_0$ at the interaction level. This approximation is consistent with the paraxial regime, in which the dynamics are dominated by the forward motion of the electron beam.
Due to the relativistic dispersion relation of the electron, the Lorentz factor cancels when the momentum is replaced by the velocity in the interaction Hamiltonian, and thus, its leading order relativistic correction vanishes.

Using the previously derived interaction Hamiltonian, we compute the scattering operator via the standard time-evolution expression
\begin{equation}
    \hat{S} = \mathcal{T} \exp \left( -\frac{\im}{\hbar} \int_{-\infty}^{+\infty} \diff t \ \hat{V}^{(\mathrm{I})}(t) \right),
\end{equation}
where $\mathcal{T}$ denotes the time-ordering operator. Since $\hat{V}_{\rm NB}^{(\mathrm{I})}(t)$ commutes with itself at all times, time ordering can be dropped. For a static magnetic moment $\vmu = \mu \vn$, the scattering operator simplifies to
\begin{equation}\label{eq:App:Scattering_NB_pre}
    \hat{S}_{\rm NB} = \exp \left( - \im \frac{r_\mathrm{e} \mu}{2 \mu_\mathrm{B}} (n_x \oy - n_y \ox) \int_{-\infty}^\infty \diff z \ \frac{F(\sqrt{|\ovx_\perp|^2 + z^2})}{(|\ovx|^2 + z^2)^{3/2}} \right).
\end{equation}
Switching to polar coordinates in the transverse plane: $(\ox,\oy) = \opr (\cos\ovarphi, \sin\ovarphi), \quad 
    (n_x,n_y) = |\vn_\perp| (\cos\varphi_\mathrm{n}, \sin\varphi_\mathrm{n})$,
where $\opr = \sqrt{\ox^2 + \oy^2}$ and $\e^{\pm\im \ovarphi} = (\ox \pm \im \oy)/|\ovx_\perp|$.
Choosing $\varphi_\mathrm{n} = 0$ without loss of generality, we find a compact form for the no-backaction scattering operator
\begin{equation}\label{eq:App:Scattering_NB}
    \hat{S}_{\rm NB}(\vartheta) = \exp \left[ - \im \vartheta |\vn_\perp| g(\opr) \sin(\ovarphi) \right],\quad \text{with} \quad \vartheta = \frac{r_\mathrm{e} \mu}{\Delta_\s \mu_\mathrm{B}}
\end{equation}
which is the dimensionless coupling constant that
sets the interaction strength, and $\Delta_\s$ defines a characteristic length scale associated with the spatial distribution of the magnetic sample, e.g., the standard deviation of a Gaussian distribution. This sets the effective size of the spin that sources the magnetic moment.
The dimensionless radial profile $g(\opr)$ is given by:
\begin{equation}\label{eq:App:general_g_func}
    g(\opr) = \frac{\Delta_\s \opr}{2} \int_{-\infty}^{+\infty} \diff z \ \frac{F(\sqrt{\opr^2 + z^2})}{(\opr^2 + z^2)^{3/2}}.
\end{equation}
which encodes the spatial profile of the interaction due to the scatterer delocalization.

Using the Jacobi-Anger expansion,
\begin{equation}
    \e^{-\im a \sin(\ovarphi)} = \sum_{k \in \mathbb{Z}} (-1)^k J_k(a) \e^{\im k \ovarphi},
\end{equation}
we write the scattering operator as
\begin{equation}
    \hat{S}_{\rm NB}(\vartheta) = \sum_{k \in \mathbb{Z}} (-1)^k J_k\left( \vartheta |\vn_\perp| g(\opr) \right) \e^{\im k \ovarphi},
\end{equation}
where $J_k$ denotes the Bessel function of the first kind.
Thus, in the absence of quantum backaction, the scattering operator acts as a coherent superposition of OAM "kicks" generated by $\e^{\im k \ovarphi}$, with amplitudes depending on the radial profile $g(\opr)$. This transforms a rotationally symmetric state into a superposition of OAM eigenstates.

To obtain an analytical expression for $g(\opr)$, we assume a radially symmetric Gaussian spin density. Using the regularizer in Eq.~\eqref{eq:App:radialF_function} and applying integration by parts, we find
\begin{align}\label{eq:App:integral_scattering_v2}
    g(\opr) 
    &= \frac{\Delta_\s \opr}{2} 
       \int_{-\infty}^{+\infty} \diff z \ 
       F\!\left( \sqrt{\opr^2 + z^2} \right) 
       \frac{\diff}{\diff z} 
       \left( \frac{z}{\opr^2 \sqrt{\opr^2 + z^2}} \right) \nonumber \\
    &= - \frac{\Delta_\s \opr}{2} 
       \int_{-\infty}^{+\infty} \diff z \ 
       \frac{z}{\opr^2 \sqrt{\opr^2 + z^2}} 
       \frac{\diff}{\diff z} 
       F\!\left( \sqrt{\opr^2 + z^2} \right) = \left(1 - \exp\!\left[-\frac{\opr^2}{2\Delta_\s^2} \right] \right) 
       \frac{\Delta_\s}{\opr},
\end{align}
where we used the identity
\begin{subequations}
\begin{align}
    \frac{\diff}{\diff z} 
    F\!\left( \sqrt{b^2 + z^2} \right) 
    = \sqrt{\frac{2}{\pi}} \, 
      \frac{z \sqrt{b^2 + z^2}}{\Delta_\s^3} 
      \exp\!\left[-\frac{b^2 + z^2}{2\Delta_\s^2}\right].
\end{align}
\end{subequations}
This result shows that the interaction strength is governed by the ratio between the radial position $\opr$ and the characteristic width $\Delta_\s$ of the spin distribution.

\medskip

\paragraph{OAM Basis Interpretation}

The structure of the scattering operator can be naturally understood using an OAM basis $\ket{r,k}$ that diagonalizes both the radius operator $\opr$ and the OAM operator $\oL_z$. These satisfy
\begin{align}
    &\opr \ket{r,k} = r \ket{r,k}, \qquad \oL_z \ket{r,k} = \hbar k \ket{r,k}, \\
    &\la r',k' \vert r,k \ra = \frac{1}{r} \delta(r'-r)\delta_{k'k}, \quad 
      \int_0^\infty \diff r \, r \sum_{k \in \mathbb{Z}} \ket{r,k} \bra{r,k} = \mathds{1}.
\end{align}
In position representation, these states are given by
\begin{equation}\label{eq:ang_Mom_state_pos}
    \la \vx_\perp \vert r,k \ra = \frac{1}{r} \delta\left( |\vx_\perp| - r \right) \frac{1}{\sqrt{2\pi}} \e^{\im k \varphi},
\end{equation}
with $\varphi$ the polar angle of $\vx_{\perp}$.
The operators $\e^{\pm \im \ovarphi} = (\ox \pm \im \oy)/\opr$ act as ladder operators in this basis, $\e^{\pm \im \ovarphi} \ket{r,k} = \ket{r,k \pm 1}$,
and are referred to as \emph{OAM kicks}.
We can now express the scattering operator in the OAM basis. Since it is diagonal in $\opr$ and decomposes into angular harmonics, we write
\begin{equation}
    \oS_{\rm NB}(\vartheta) = \sum_{l,k \in \mathbb{Z}} \int_0^\infty \diff r \, r \, (-1)^k J_k\left( \vartheta |\vn_\perp| g(r) \right) \ket{r,k+l} \bra{r,l}.
\end{equation}
This shows how the scattering operator selectively modifies the OAM state based on the radial wavefunction profile.

Note that the scattering operator $\oS_{\rm NB}(\vartheta)$ commutes with the longitudinal boost phase from the SVEA ansatz,
\begin{equation}
    \left[ \oS_{\rm NB}(\vartheta), \exp \left( \im \frac{\vp_0 \cdot \ovx - E_0 t}{\hbar} \right) \right] = 0,
\end{equation}
which implies that $\oS_{\rm NB}(\vartheta)$ can be consistently applied to both the boosted wavefunction $\ket{\psi(t)}$ and its envelope $\ket{u(t)}$.
The OAM basis helps to understand how the scattering operator acts and how to construct optimal measurements for sensing.

\subsection{Scattering Operator with Spin Backaction}
\label{App:Met:ScattBa}

We begin with the full backaction Hamiltonian $\oH_{\rm BA}$, which is defined from Eq.~\eqref{eq:Met:RegHam} with $\vmu=\mu\ovsigma$, as follows
\begin{subequations}\label{eq:Met:RegHam_BA}
    \begin{align}
        \oH_\mathrm{BA} =&  \sqrt{(mc^2)^2+c^2\left| \ovp\right|^2} + \hbar \omega_0 (\vn_0 \cdot \ovsigma) + \oV_{\rm BA} \\
        \oV_{\rm BA} :=& \frac{\hbar r_\mathrm{e}}{2m} \left( \frac{\mu}{\mu_\mathrm{B}} \right) \ovsigma \cdot \left( \frac{F(|\ovx|)}{|\ovx|^3} \frac{\ovx\times \ovp}{\gamma(|\ovp|)} \right),
    \end{align}
\end{subequations}
which incorporates the quantum spin dynamics of the magnetic sample:
Using the same boosted wavefunction ansatz, as in Eq.~\eqref{eq:App:boosted_wavefucntion}, the effective Hamiltonian governing the envelope state becomes
\begin{equation}
    \oH_\mathrm{BA}^{\mathrm{eff}} = \underbrace{ v_0 \op_z+ \frac{\ovp^2}{2\gamma(|\vp_0|)m} + \frac{\hbar \omega_0}{2} \left( \vn_0 \cdot \ovsigma \right)}_{=\oH_{\rm BA}^{(0)}}
    + \underbrace{\frac{\hbar r_\mathrm{e}}{2} \left( \frac{\mu}{\mu_\mathrm{B}}\right) v_0 \left( F(|\ovx|) \frac{\ovsigma \cdot \left( \ovx \times \ve_z \right)}{|\ovx|^3} \right)}_{=\oV_{\rm BA}},
\end{equation}
where the relativistic corrections are cancelled out as shown in Appendix~\ref{App:Met:ScattNoBa}.
In the interaction picture and under the paraxial regime the propagation is mainly given by the longitudinal component of the electron beam along $z$. Then the interaction term becomes
\begin{align}
    \oV^{(\mathrm{I})}_{\rm BA}(t) &= \e^{\im \oH_{\rm BA}^{(0)} t/\hbar} \oV_{\rm BA} \e^{-\im \oH_{\rm BA}^{(0)} t/\hbar} \approx \frac{\hbar r_\mathrm{e}}{2} \left( \frac{\mu}{\mu_\mathrm{B}} \right) v_0 \ovsigma \cdot 
    \frac{F(|\ovx + v_0 t \ve_z|) R(\vn_0,-\omega_0 t)}{|\ovx + v_0 t \ve_z|^3}
    \begin{pmatrix}
         \oy \\ -\ox \\ 0
    \end{pmatrix},
\end{align}
with $R(\vn_0, \omega_0 t)$ the rotation about $\vn_0$ by angle $\omega_0 t$. This arises from spin precession under $\oH_{\rm BA}^{(0)}$. Since $\oV^{(\mathrm{I})}_{\rm BA}(t)$ commutes with itself at all times, the scattering operator can be written as
\begin{equation}
    \oS_{\rm BA} = \exp \left[ - \im \frac{r_\mathrm{e}}{2} \left( \frac{\mu}{\mu_\mathrm{B}} \right) \ovsigma \cdot \int_{-\infty}^{+\infty} v_0 \diff t \ \frac{F(|\ovx + v_0 t \ve_z|) R(\vn_0,-\omega_0 t)}{|\ovx + v_0 t \ve_z|^3}
    \begin{pmatrix}
         \oy \\ -\ox \\ 0
    \end{pmatrix} \right].
\end{equation}
and depends on the temporal integral
\begin{align}
    \int_{-\infty}^{+\infty} v_0 \diff t \ 
    \frac{F(|\ovx + v_0 t \ve_z|) R(\vn_0, -\omega_0 t)}{|\ovx + v_0 t \ve_z|^3}
    &= \int_{-\infty}^{+\infty} \diff z \ 
    \frac{F\left(|\ovx_\perp+z \ve_z| \right) R\left( \vn_0, \frac{\omega_0}{v_0}(\oz-z) \right)}{|\ovx_\perp+z \ve_z|^3} \nonumber \\ 
    &= R\left( \vn_0, \frac{\omega_0}{v_0}\oz \right) 
    \int_{-\infty}^{+\infty} \diff z \ 
    \frac{F\left(|\ovx_\perp+z \ve_z| \right) R\left( \vn_0, \frac{\omega_0}{v_0}(-z) \right)}{|\ovx_\perp+z \ve_z|^3}.
\end{align}
where we first split the position operator into a component parallel to the beam axis and a transverse part. The longitudinal operator is then decomposed in its spectrum $\int_{-\infty}^{+\infty}\diff \tilde{z}\ket{\Tilde{z}}\bra{\Tilde{z}}$, and by performing the substitution $z=\Tilde{z}+v_0 t$ we isolate the free longitudinal propagation of the electron. This step allows us to factor out the rotation matrix, which considerably simplifies the expression in the interaction picture.

However, the conditions on the longitudinal and transversal widths, $\Delta_{\e, \|}\ll \frac{v_0}{\omega_0}$ and $
\Delta_{\e} \ll \frac{v_0}{\omega_0},$
are satisfied for a sufficiently small bias field. Under these conditions, the rotation matrices vary negligibly over the support of the free-electron wavefunction, meaning the electron effectively probes a constant spin orientation. Consequently, we approximate $R\left(\vn_0, \omega_0 \oz/v_0 \right) \approx \mathds{1},$ $R\left( \vn_0, -\omega_0 z/v_0 \right) \approx \mathds{1}$, where the first approximation follows from the longitudinal width condition, and the second from the transverse width condition.
With this, the integral matches the integral in Eq.~\eqref{eq:App:Scattering_NB_pre} for the no-backaction case (see Appendix~\ref{App:Met:ScattNoBa}), and the scattering operator reduces to:
\begin{align}
    \oS_{\rm BA}(\vartheta) &= \exp \left( -\im \vartheta g\left( \opr \right) \left[ \osigma_x \sin(\ovarphi) - \osigma_y \cos(\ovarphi) \right] \right).
\end{align}
where $g\left( \opr \right)$ is defined as in Eq.~\eqref{eq:App:general_g_func}.
The angular dependence can be compactly expressed as 
$\osigma_x \sin(\ovarphi) - \osigma_y \cos(\ovarphi) = \im \left( \osigma_+ \e^{-\im \ovarphi} - \osigma_- \e^{\im \ovarphi} \right)$, so that the scattering operator reads
\begin{equation}\label{eq:Sop_Back_origin}
    \oS_{\rm BA}(\vartheta) = \exp \Big( \vartheta g(\opr) \left[ \osigma_+ \e^{-\im \ovarphi} - \osigma_- \e^{\im \ovarphi} \right] \Big) = \oK_0 + \left( \oK_- \osigma_+ - \oK_+ \osigma_- \right).
\end{equation}
Here, we make use of $\left[ \e^{-i \ovarphi} \hat{\sigma}_+ - \e^{i \ovarphi} \hat{\sigma}_- \right]^2 = -\{ \osigma_+, \osigma_- \} = - \mathds{1}$, and we define $\oK_0 = \cos[\vartheta g(\opr)]$ and $\oK_\pm = \sin[\vartheta g(\opr)] \e^{\pm\im \ovarphi}$ as in \eqref{eq:Kops}.  
This decomposition shows that spin flips ($\osigma_\pm$) are strictly accompanied by $\pm \hbar$ changes in orbital angular momentum. Unlike the no-backaction case, where the Jacobi-Anger expansion allows arbitrary OAM exchange, the backaction-limited interaction enforces quantized angular momentum transfer through spin-orbit coupling.

\paragraph{Derivation of the Scattering Channel}
\label{App:Met:ScattChannel}

From the Scattering operator in Eq.~\eqref{eq:Sop_Back_origin} we calculate the evolution of the electron's motional state after interacting with a magnetic sample. This channel is obtained by tracing out the spin degree of freedom following the unitary scattering transformation $\oS_{\rm BA}(\vartheta)$ applied to the joint spin-motional state.
The effective channel $\Lambda_{\vartheta, \vc}$ acts on the initial motional state $\orho_\mathrm{e}(0)$ and depends on the dimensionless interaction strength $\vartheta$ and the Bloch vector $\vc$ that characterizes the initial state of the sample spin:
\begin{equation}
    \orho_{\rm BA}(\vartheta)=\Lambda_{\vc, \vartheta} \left[ \orho_\mathrm{e}(0) \right]  = \tr_\mathrm{S} \left\{ \oS_{\rm BA}(\vartheta) \left( \orho_\mathrm{e}(0) \otimes \orho_\mathrm{S} \right) \oS^\dagger_{\rm BA}(\vartheta) \right\},
\end{equation}
where $ \orho_\mathrm{S} = \frac{1}{2} \left( \mathds{1} + \vc \cdot \ovsigma \right),$ with $\quad |\vc| \le 1$.
Substituting the scattering operator in Eq.~\eqref{eq:Sop_Back_origin} into the channel definition and performing the partial trace over the spin subsystem yields:
\begin{align}
    \label{eq:App:ScattChanSimplified}
    \orho_{\rm BA}(\vartheta)=\Lambda_{\vc,\vartheta} \left[ \orho_\mathrm{e}(0) \right] 
    &=\oK_0 \orho_\mathrm{e}(0) \oK_0^\dagger
    +\frac{1}{2} \left(\oK_+ \orho_\mathrm{e}(0) \oK_+^\dagger + \oK_- \orho_\mathrm{e}(0) \oK_-^\dagger \right)
    + \frac{c_z}{2} \left( \oK_+ \orho_\mathrm{e}(0) \oK_+^\dagger - \oK_- \orho_\mathrm{e}(0) \oK_-^\dagger \right) \nonumber \\ 
    &\quad + \im |\vc_\perp| \left( \oK_0 \orho_\mathrm{e}(0) \sin(\vartheta g(\opr)) \sin(\ovarphi-\varphi_c) - \sin(\vartheta g(\opr)) \sin(\ovarphi-\varphi_c) \orho_\mathrm{e}(0) \oK_0  \right),
\end{align}
where $c_x + \im c_y = |\vc_\perp| \e^{\im \varphi_c}$.
We emphasize that the phase $\varphi_c$ can be set to zero without loss of generality by choosing the coordinate system in the transverse plane appropriately.

This scattering channel forms the basis for evaluating measurement statistics. For instance, the likelihood function for a position measurement is unaffected by the scattering process and reads:
\begin{align}\label{eq:App:BA:ProbPos}
    P_{\rm BA}(\vx_\perp \vert \vartheta) &= \bra{\vx_\perp} \Lambda_{\vc,\vartheta} \left[ \orho_\mathrm{BA}(0) \right] \ket{\vx_\perp} 
    = \bra{\vx_\perp} \orho_\mathrm{BA}(0) \ket{\vx_\perp} = P(\vx_\perp \vert 0),
\end{align}
independent of $\vartheta$ and $\vc$.
For momentum measurements, assuming the initial motional state is pure, $\orho_\mathrm{e}(0) = \ket{\psi(0)} \bra{\psi(0)}$, the likelihood becomes:
\begin{align}
    \label{eq:App:BA:ProbMom}
    P_{\rm BA}(\vp_\perp \vert \vartheta) &= |\bra{\vp_\perp} \oK_0 \ket{\psi(0)}|^2 
    + \frac{1+c_z}{2} |\bra{\vp_\perp} \oK_+ \ket{\psi(0)}|^2 
    + \frac{1-c_z}{2} |\bra{\vp_\perp} \oK_- \ket{\psi(0)}|^2 \nonumber \\
    &\quad + 2 |\vc_\perp| \Im \left\{ \bra{\vp_\perp} \sin(\vartheta g(\opr)) \sin(\ovarphi) \ket{\psi(0)} \bra{\psi(0)} \oK_0 \ket{\vp_\perp} \right\}.
\end{align}
We will use this result in subsequent sections to compute the classical Fisher information (CFI) associated with OAM and momentum-based measurements.

\section{Fisher information for magnetic moment estimation}\label{App:QFI}

In the no-backaction case with a pure zero-OAM electron state, we find that the CFI for momentum measurements saturates the QFI. This is due to the following general argument \cite{wasak_optimal_2016}: Consider a measurement basis $(\ket{\xi})$ and a quantum state $\ket{\psi (\vartheta)}$, which we can expand in said basis as
\begin{equation}
    \ket{\psi(\vartheta)} = \int \diff \xi \ \psi(\xi \vert \vartheta) \ket{\xi} 
    = \int \diff \xi \ \sqrt{p(\xi \vert \vartheta)} \e^{\im \phi(\xi\vert \vartheta)} \ket{\xi}.
\end{equation}
Inserting this into the pure-state QFI expression \eqref{eq:QFI_pure}, we obtain a relation between this QFI and the CFI associated to the likelihood of measurement outcomes $p(\xi|\vartheta)$, 
\begin{equation}
    \label{eq:OptCrit}
    \cJ(\vartheta) = \cI(\vartheta) + 4 \mathbb{V}_\vartheta \left[ \frac{\partial \phi}{\partial \vartheta}  \right], \qquad \text{where} \quad \mathbb{V}_\vartheta \left[ \frac{\partial \phi}{\partial \vartheta}  \right] = \int\diff\xi \, p(\xi|\vartheta) \left[ \frac{\partial \phi(\xi|\vartheta)}{\partial \vartheta}  \right]^2 - \left[ \int\diff\xi \, p(\xi|\vartheta) \frac{\partial \phi(\xi|\vartheta)}{\partial \vartheta} \right]^2.
\end{equation}
The QFI and the CFI coincide whenever the variance expression vanishes. In particular, this is the case when the wavefunction $\psi(\xi|\vartheta)$ is real-valued for all $\xi$. We now prove that this holds for any $\vartheta$ in the momentum basis. The only assumption we need is that the initial $\ket{\psi(0)}$ is a zero-OAM state, i.e., $\la \vx_\perp \vert \psi(0) \ra$ only depends on $r=|\vx_{\perp}|$ and not the polar angle $\varphi (\vx_{\perp})$. Applying the scattering operator \eqref{eq:Met:ScattNoBa} and expanding the exponential in a power series, the scattered wavefunction in position representation becomes
\begin{equation}
    \la \vx_{\perp}|\psi(\vartheta)\ra =  \sum_{n=0}^\infty \frac{(-i\vartheta)^n}{n!}  g^n(r)\sin^n (\varphi) \la \vx_\perp | \psi(0) \ra = \sum_{n=0}^\infty \frac{\vartheta^n}{n!} \left[\frac{-i g \left(\sqrt{x^2+y^2}\right) y}{\sqrt{x^2+y^2}}\right]^n \la x,y | \psi(0) \ra .
\end{equation}
For even $n$, the summand is an even, real-valued function of $\vx_{\perp}$, and thus its Fourier transform is real-valued. For  odd $n$, on the other hand, the summand is even in $x$ but odd in $y$. Given the imaginary prefactor, $(-\im)^n = \pm \im$, the Fourier transform is also real-valued. In total, the Fourier transform of $\la \vx_{\perp}|\psi(\vartheta)\ra$, which gives the scattered state in momentum representation, is real-valued, which proves our claim. 

\begin{figure}
    \centering
    \includegraphics[width=0.75\linewidth]{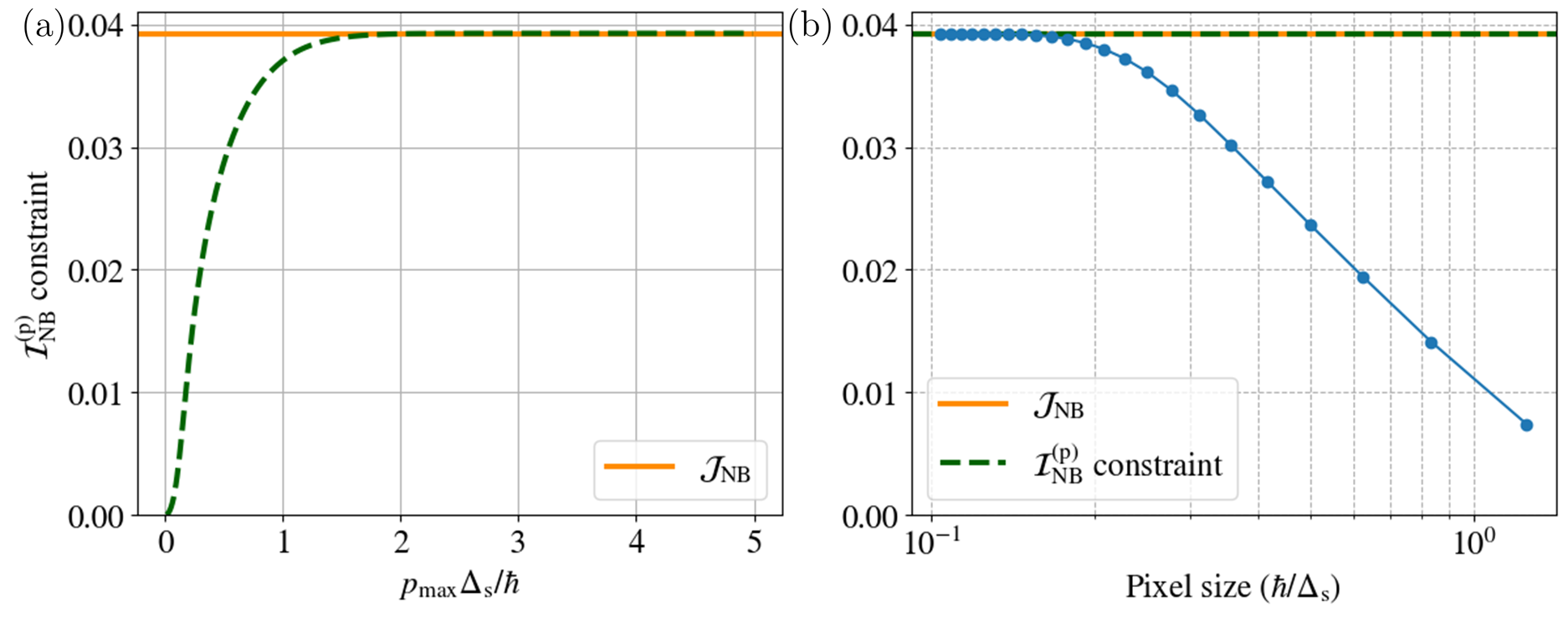}
    \caption{(a) CFI from a momentum measurement with $\Delta_\e = 10~\mathrm{nm}$ and $\Delta_\s = 1~\mathrm{nm}$, 
restricted to momentum values $p \leq p_{\rm max}$ with sufficient resolution, shown as a function of $p_{\rm max}$. The quantum limit, given by $\mathcal{J}_{\rm NB}$, is indicated by the orange line. 
(b) CFI from a momentum measurement with $p_{\rm max} = 5\hbar/\Delta_\s$, plotted as a function of pixel size (blue curve with dots). For small pixel sizes, the result converges to the constrained CFI from a momentum 
measurement (green dashed curve). }
    \label{fig:cfi_momentum_constraint}
\end{figure}

An important detail for diffraction mode imaging is that the characteristic length scales of the probe wavefunction and the (smaller) scatterer determine the size and features of the imaged momentum-space wavefunction. The detector must be sufficiently large and its pixels sufficiently small to capture those features. As illustrated in Fig.~\ref{fig:cfi_momentum_constraint}, the detector's captured momentum range must extend beyond the inverse length scale of the scatterer, while the pixel size must be smaller than the inverse size of the wavefunction. When these conditions are met, the QFI can be saturated in practice.

For the backaction case in which the magnetic moment is sourced by a single spin, diffraction-mode measurements no longer saturate the QFI. We show this by deriving the corresponding CFI in the weak-sample limit, $\vartheta \to 0$. From the expression \eqref{eq:finalstate_BA} for the scattered state, we can deduce its derivative with respect to $\vartheta$ at zero, $\orho_{\e}' (0) = -\im |\vc_{\perp}| [g(\opr)\sin\ovarphi,\orho_{\e}(0)]$. For the likelihood of measurement outcomes, $P(\vp_{\perp}|\vartheta) = \bra{\vp_{\perp}} \orho_{\e} (\vartheta) \ket{\vp_{\perp}}$, we then get $P(\vp_{\perp}|0) = |\la \vp_\perp \vert \psi(0) \ra \vert^2$ and
\begin{align}
    \frac{\partial P(\vp_\perp \vert 0)}{\partial \vartheta}   &= -2 |\vc_\perp| \ \Im \left\{ \bra{\psi(0)} g(\opr)\sin(\ovarphi) \ket{\vp_\perp} \la \vp_\perp \vert \psi(0) \ra \right\} = 2i |\vc_\perp| \bra{\psi(0)} g(\opr)\sin(\ovarphi) \ket{\vp_\perp} \la \vp_\perp \vert \psi(0) \ra .
\end{align}
In the second step, we make use of the fact that the initial radially symmetric $\la \vx_{\perp}|\psi (0)\ra$ depends only on $r=|\vx_{\perp}|$. It is thus an even function in $\vx_{\perp} = (x,y)$, and its Fourier transform $\la \vp_{\perp}|\psi (0)\ra$ is real-valued. On the other hand, $\bra{\vx_{\perp}}g(\opr) \sin(\ovarphi) \ket{\psi_0}$ is even in $x$ and odd in $y$, and so its Fourier transform purely imaginary. The CFI at $\vartheta=0$ becomes, 
\begin{align}
    \cI_\mathrm{BA}^{\rm (p)}(0) &= 4 |\vc_\perp|^2 \int \diff^2 \vp_\perp \ \frac{|\bra{\psi(0)} g(\opr)\sin(\ovarphi) \ket{\vp_\perp}|^2 |\la \vp_\perp \vert \psi(0) \ra|^2}{|\la \vp_\perp \vert \psi(0) \ra|^2} = 4 |\vc_\perp|^2 \la \psi(0)|g^2(\opr) \sin^2(\ovarphi)|\psi(0)\ra = |\vc_\perp|^2 \cG.
\end{align}
From the symmetry of the scattering transformation with respect to $\vc \to -\vc$, we infer that the CFI must be an even function in $\vartheta$, Hence, for small $\vartheta \neq 0$, the CFI deviation should be of second order, $\cI_\mathrm{BA}^{\rm (p)}(\vartheta) = \cI_\mathrm{BA}^{\rm (p)}(0) + \cO (\vartheta^2)$.

\section{Discrimination without backaction}

\subsection{Quantum trace distance and optimal POVM}\label{App:QTraceDerivation}

The quantum trace distance between the unscattered electron state $\ket{\psi(0)}$ and the scattered state $\ket{\psi(\vartheta)}$ provides a fundamental bound on the success probability for discriminating the presence or absence of interaction. We evaluate this using the Jacobi-Anger expansion of the scattering operator from Eq.~\eqref{eq:Met:ScattNoBa}.
Assuming a radially symmetric initial state $\ket{\psi(0)}$, the scalar product between the two states reduces to $\braket{\psi(0)}{\psi(\vartheta)} 
= \langle J_0 \left( \vartheta |\vn_\perp| g(\opr) \right)\rangle_0$, which results in the quantum trace distance, 
\begin{equation}
    D^\mathrm{Q} \left( \ket{\psi(0)}, \ket{\psi(\vartheta)} \right) 
    = \sqrt{1 - \left| \braket{\psi(0)}{\psi(\vartheta)} \right|^2}= \sqrt{1 - \left\langle J_0 \left( \vartheta |\vn_\perp| g(\opr) \right) \right\rangle_0^2}.
\end{equation}
For small interaction parameters $\vartheta \ll 1$, we expand the Bessel function as $J_0(x) = 1 - \frac{x^2}{4} + \mathcal{O}(x^4),$ and find
\begin{align}
    \braket{\psi(0)}{\psi(\vartheta)} 
    &= 1 - \frac{1}{4} \vartheta^2 |\vn_\perp|^2 \langle g(\opr)^2 \rangle_0 + \mathcal{O}(\vartheta^4).
\end{align}
Thus, the trace distance becomes
\begin{align}
    D^\mathrm{Q} \left( \ket{\psi(0)}, \ket{\psi(\vartheta)} \right) 
    &= \frac{\vartheta}{\sqrt{2}} \, |\vn_\perp| \sqrt{\langle g(\opr)^2 \rangle_0} + \mathcal{O}(\vartheta^3) = \frac{\vartheta}{2} |\vn_\perp|\sqrt{\cG} + \mathcal{O}(\vartheta^3).
\end{align}

The optimal binary measurement that achieves this trace distance corresponds to projection onto the positive eigenspace of the operator $\ket{\psi(\vartheta)}\bra{\psi(\vartheta)} - \ket{\psi(0)}\bra{\psi(0)}$ \cite{nielsen2010quantum}. Let us first construct a two-dimensional orthonormal basis in the span of the two states,
\begin{equation}
    \ket{0} = \ket{\psi(0)}, \qquad
    \ket{\vartheta} = \frac{\ket{\psi(\vartheta)} - \langle \psi(0) | \psi(\vartheta) \rangle \ket{\psi(0)}}{D^\mathrm{Q}}=\frac{1}{D^\mathrm{Q}} \left( \ket{\psi(\vartheta)} - \sqrt{1 - (D^\mathrm{Q})^2} \ket{\psi(0)} \right).
\end{equation}
Here, we have made use of the fact that, for radially symmetric initial states, the overlap $\langle \psi(0) | \psi(\vartheta) \rangle$ is real-valued. In the so defined orthonormal basis $\{ \ket{\vartheta}, \ket{0} \}$, the above operator becomes
\begin{equation}
    \ket{\psi(\vartheta)}\bra{\psi(\vartheta)} - \ket{\psi(0)}\bra{\psi(0)} = D^\mathrm{Q}
    \begin{pmatrix}
        D^\mathrm{Q} & \sqrt{1 - (D^\mathrm{Q})^2} \\
        \sqrt{1 - (D^\mathrm{Q})^2} & -D^\mathrm{Q}
    \end{pmatrix}.
\end{equation}
Diagonalizing yields the positive eigenvector:
\begin{equation}
    \ket{+} = \sqrt{\frac{1 + D^\mathrm{Q}}{2}} \ket{\vartheta} 
    + \sqrt{\frac{1 - D^\mathrm{Q}}{2}} \ket{0} = \frac{1}{D^\mathrm{Q}} \left( \sqrt{\frac{1 + D^\mathrm{Q}}{2}} \ket{\psi(\vartheta)} 
    - \sqrt{\frac{1 - D^\mathrm{Q}}{2}} \ket{\psi(0)} \right).
\end{equation}
Hence, the optimal measurement for distinguishing $\ket{\psi(\vartheta)}$ from $\ket{\psi(0)}$ is given by the projective POVM $\left\{ \ket{+}\bra{+}, \mathds{1} - \ket{+}\bra{+} \right\}$. While this strategy saturates the theoretical bound, its practical implementation may not be feasible in realistic experimental conditions, where full knowledge of the states is unavailable. This motivates our investigation of more accessible and standard measurement schemes.

\subsection{First order perturbative defocus planes imaging}\label{App:NB:First_order_defocus}

Here we derive perturbative results for the scattering amplitudes on an arbitrary defocus plane in TEM imaging from image mode to diffraction mode, to first order in $\vartheta$. To this end, we model the propagation of the electron beam through the imaging objective lens to the detector plane by the following three-step transformation of the transverse electron state: First, the electron wave propagates a distance $z$ along the beam axis, which amounts to an effective free propagation time $t=z/v_0$ and the application of the unitary 
\begin{equation}
    \oU_\mathrm{free} (z) = \exp \left( -\im \frac{\vp^2_\perp}{2\gamma(|\vp_0|)m \hbar} t  \right) 
    = \exp \left( -\im \frac{z}{2p_0} \frac{\vp_\perp^2}{\hbar} \right).
\end{equation}
Second, the electron passes an ideal thin lens of focal length $f$ without aberrations or apertures, which imprints a quadratic phase onto the transverse wavefunction, described by the unitary (with $k_0 = p_0/\hbar=\gamma(|\vp_0|)mv_0/\hbar$)
\begin{equation}
    \oU_\mathrm{lens} = \exp \left( - \im \frac{k_0}{2f} \ovx_\perp^2 \right).
\end{equation}
Third, the electron propagates another distance $z$ to the detector. The overall unitary transformation that concatenates the three steps can be rewritten in terms of auxiliary harmonic oscillator quadratures~\cite{Quijas_2007},
\begin{equation}
    \label{eq:Met:OpticUnitary}
    \oU_\mathrm{opt}(z) = \oU_\mathrm{free} (z) \oU_\mathrm{lens} \oU_\mathrm{free} (z) = \e^{-\im \alpha(z) \left( \hat{a}^\dagger \hat{a} + \hat{b}^\dagger \hat{b} + 1 \right)}, \qquad \text{with}\quad \alpha(z)=\arccos\left(1-\frac{z}{f}\right) .
\end{equation}
A position measurement of the propagated electron on the detector plane is equivalent to measuring the electron state immediately after scattering in the back-propagated basis $\ket{\vxi_z}=\hat{U}^\dagger_{\rm opt}(z)\ket{\vx_\perp}$, where $\vp_\perp=\hbar\kappa^2\vxi_f$ and $\kappa=~\sqrt{k_0}\left(z(2f-~z)\right)^{-1/4}$.

In Eq.~\eqref{eq:Met:OpticUnitary}, the mode operators $\oa,\ob$ define a two-dimensional isotropic harmonic oscillator. Its eigenfunctions in polar coordinate representation~\cite{enz2000wave}, $\vx_{\perp} = (r,\varphi)$, are given by
\begin{subequations}
    \begin{align}
\bra{n, l} \vx_\perp\ra =&\cR_{n,l}(r)\e^{il \varphi},\\
\cR_{n,l}(r)=&N_{n,l}r^{|l|}\exp{-\tfrac{1}{2}\kappa^2r^2}L_n^{|l|}(\kappa^2r^2), 
    \quad \text{with }\ N_{n,l}=\sqrt{\frac{n! \kappa^{2{|l|}+2}}{\pi(n+{|l|})!}},
    \end{align}
\end{subequations}
and their corresponding eigenvalues are $2n + |l| + 1$.
The $L_n^{|l|}(x)$ are associated Laguerre polynomials with $n\geq 0$ and $l\in\mathbb{Z}$. The scattering amplitudes in this basis are
\begin{equation}\label{eq:amplitude_HOScatter}
\la \vxi_z | \hat{S}_{\rm NB}\ket{\psi(0)}=\int \diff^2x'_\perp \bra{\vx_\perp} \oU_{\rm opt}(z)\ket{\vx'_\perp}\bra{\vx'_\perp} \hat{S}_{\rm NB}\ket{\psi(0)}
\end{equation}
The first matrix element can be expanded as
\begin{equation}
    \bra{\vx_\perp} \oU_{\rm opt}(z)\ket{\vx'_\perp}=\sum_{n\geq0,l\in \mathbb{Z}}\bra{\vx_\perp} n , l\ra \e^{-\im \alpha (2n+|l|+1)}\bra{n, l}\vx'_\perp\ra.
\end{equation}
The scattering operator in Eq.~\eqref{eq:App:Scattering_NB}, to first order in $\vartheta$, can be separated into two leading contributions, 
\begin{equation}
    \oS_{\rm NB}=\mathbb{1}+\oS_{\rm NB}^{(1)}(\ovx_\perp)+\cO(\vartheta^2),\quad 
\text{with} \quad
    \oS_{\rm NB}^{(1)}(\ovx_\perp)=-\im\vartheta |\vn_\perp| g(\opr)\sin(\ovarphi),
\end{equation}
with $(r,\varphi)$ the polar coordinate representation of $\vx_{\perp}$. 
Substituting this into Eq.~\eqref{eq:amplitude_HOScatter}, we obtain
\begin{equation}
    \begin{aligned}
\la \vxi_z | \hat{S}_{\rm NB}(\vartheta)\ket{\psi(0)}
=& \sum_{n\geq0,l\in \mathbb{Z}}\bra{\vx_\perp} n , l\ra \e^{-\im \alpha(z) (2n+|l|+1)} {c}_{n, l}(\kappa, \Delta_\e, \Delta_\s)+\cO(\vartheta^2), \quad \text{where}\\
{c}_{n, l}(\kappa, \Delta_\e, \Delta_\s )=&\int \diff^2x'_\perp \bra{n, l}\vx'_\perp\ra \Big(1+\tilde{S}_{\rm NB}^{(1)}(r')\sin\varphi\Big)\psi_0(r')
    \end{aligned}
\end{equation}
and $\tilde{S}_{\rm NB}^{(1)}(r)=-\im\vartheta |\vn_\perp| g(r)$.

The coefficients ${c}_{n, l}$ are nonzero only for specific OAM values, as dictated by the azimuthal integral,
\begin{equation}\label{eq:App:general_cnl_coef}
    \begin{aligned}
{c}_{n, l}(\kappa, \Delta_\e, \Delta_\s)=&\int_0^\infty \diff r'~r' \int_0^{2\pi}\diff \varphi \cR_{n, l}(r') \e^{il\varphi}\Big(1+\tilde{S}_{\rm NB}^{(1)}(r')\sin(\varphi)\Big)\psi_0(r')\\
=&\int_0^\infty \diff r'~r' \cR_{n, l}(r') \Big(2\pi \delta_{l,0}+i\pi l \tilde{S}_{\rm NB}^{(1)}(r')\delta_{l,\pm 1}\Big)\psi_0(r').
    \end{aligned}
\end{equation}
Indeed, when an incident wavepacket with zero OAM interacts with the magnetic dipole (spin), the unique states of OAM with non-zero amplitude are $l=0$ from the unscattered state and $l=\pm 1$ from the scattered state. Then, the probability amplitudes correspond to
\begin{equation}
    \la \vxi_z | \hat{S}_{\rm NB}(\vartheta)\ket{\psi(0)}
=\sum_{n=0}^\infty\sum_{l=-1}^1\bra{\vx_\perp} n , l\ra \e^{-\im \alpha(z) (2n+|l|+1)}{c}_{n, l}(\kappa, \Delta_\e, \Delta_\s)+\cO(\vartheta^2)
\end{equation}
and each case the ${c}_{n, l}$ coefficients have the following form:
\begin{subequations}
    \begin{align}
c_{n, 0}(\kappa, \Delta_\e, \Delta_\s)=&2\pi\int_0^\infty \diff r'~r' \cR_{n,0}(r') \psi_0(r')\\
c_{n, \pm 1}(\kappa, \Delta_\e, \Delta_\s)=&\pm \vartheta |\vn_\perp|\tilde{c}_n(\kappa, \Delta_\e, \Delta_\s)\\
\tilde{c}_n(\kappa, \Delta_\e, \Delta_\s)=&\pi \int_0^\infty \diff r'~r' \cR_{n,1}(r') g(r') \psi_0(r')
    \end{align}
\end{subequations}
which can be calculated for any radial symmetric spin smearing function and electron beam wavefunction.

Equivalently, the probability amplitudes that we aim to calculate are given by:
\begin{equation}
\la \vxi_z | \hat{S}_{\rm NB}(\vartheta)\ket{\psi(0)}= \la \vxi_z |\psi(0)\ra+\vartheta \la \vxi_z |\psi^\prime(0)\ra+\cO(\vartheta^2), 
\end{equation}
in which $\ket{\psi^\prime(0)}=-\im |\vn_\perp|g(\opr)\sin(\ovarphi)\ket{\psi(0)}$, and $\vxi_z=(r, \varphi)$ is a set of transverse coordinates at a certian defocus $z$. The non-scattered and the scattered part of the wavefunction in the $\vxi_z$ coordinate representation are given respectively by
\begin{subequations}
    \begin{align}
\la \vxi_z |\psi(0)\ra=& \sum_{n=0}^\infty \e^{-\im \alpha(z)(2n+1)}\cR_{n,0}(r) {c}_{n, 0}(\kappa, \Delta_\e, \Delta_\s),\label{eq:psi0_in_defocus_plane_general}\\
\la \vxi_z | \psi^\prime(0)\ra=& |\vec{n}_\perp|\sin(\varphi)~\Xi_z(r)\label{eq:scattpsi0_in_defocus_plane}
    \end{align}
\end{subequations}
which has a general $\sin(\varphi)$ azimuthal angle dependence at any defocused plane. Additionally, we define
\begin{equation}\label{eq:App:generalXi}
    \begin{aligned}
\Xi_z(r)=&-2 \im \sum_{n=0}^\infty \e^{-2\im \alpha(z)(n+1)}\cR_{n,1}(r)\tilde{c}_{n}(\kappa, \Delta_\e, \Delta_\s)\\
=&-2 \im \sum_{n=0}^\infty \e^{-2\im \alpha(z)(n+1)}\cR_{n,1}(r)  \int_0^\infty \diff r'~r' \cR_{n,1}(r') g(r') \psi_0(r'),
    \end{aligned}
\end{equation}
which determines the radial dependence of the probability amplitude of the scattered part of the wavefunction at a given defocus.

We now evaluate the coefficients $c_{n,l}$ for the case of a Gaussian incident electron beam wavefunction. We shall make use of the following integral identities \cite{Zwillinger2014}:
\begin{equation}
    \int_0^\infty \diff u \, \exp{-b u} L_n^0(u)=\frac{1}{b}\left(1-\frac{1}{b}\right)^n, \qquad  \int_0^\infty \diff u \, \exp{-b u} L_n^1(u)=1-\left(1-\frac{1}{b}\right)^{n+1}.
\end{equation}
For $l=0$, the coefficient $c_{n,0}$ is given by
\begin{equation}
    \begin{aligned}
{c}_{n, 0}(\kappa, \Delta_\e, \Delta_\s)=& \int_0^\infty \diff r'~r' \,
N_{n,0}\exp{-\tfrac{1}{2}\kappa^2 r^{'2}}L_n^0(\kappa^2 r^{'2})\frac{2\pi}{\sqrt{2\pi}\Delta_\e}\exp{-\tfrac{r^{'2}}{4\Delta_\e^2}}\\
=& N_{n,0}\frac{\sqrt{2\pi}}{\Delta_\e} \int_0^\infty \frac{\diff u}{2\kappa^2}\exp{-\frac{\kappa^{'2}}{2\kappa^2}u}L_n^0(u) = \frac{\sqrt{2}}{\Delta_\e}\frac{\kappa}{\kappa^{'2}} \left(1-\frac{2\kappa^2}{\kappa^{'2}}\right)^n, \quad \text{where}\ \ \kappa'^2=\kappa^2+\frac{1}{2\Delta_\e^2}.
    \end{aligned}
\end{equation}
For the coefficients with $l=\pm 1$, we employ Eq.~\eqref{eq:App:general_cnl_coef} and insert Eq.~\eqref{eq:App:integral_scattering_v2} for $g(r)$ in the Gaussian model,
\begin{equation}
    \begin{aligned}
\tilde{c}_{n}(\kappa, \Delta_\e, \Delta_\s)=&\sqrt{\frac{\pi}{2}}\frac{\Delta_\s }{\Delta_\e}N_{n, 1}\int_0^\infty \diff r' \, r' L_n^1(\kappa^2 r^{'2})
\Big(\exp{-\tfrac{1}{2}\kappa^{'2} r^{'2}}-\exp{-\tfrac{1}{2}\bar{\kappa}^{2} r^{'2}}\Big)\\
=&\sqrt{\frac{1}{2(n+1)}}\frac{ \Delta_\s}{2\Delta_\e} 
\left[\left(1-\frac{2\kappa^2}{\bar{\kappa}^{2}}\right)^{n+1}-\left(1-\frac{2\kappa^2}{\kappa^{'2}}\right)^{n+1}\right], \qquad \text{where}\ \ \bar{\kappa}^2=\kappa^{\prime 2}+\frac{1}{\Delta_\s^2}.
    \end{aligned}
\end{equation}

Next, we make use of the generating function of Laguerre polynomials,
\begin{subequations}
    \begin{align}
\sum_{n=0}^\infty L_n(a) b^n \e^{-2\im \alpha n}=\frac{1}{1-be^{-2\im \alpha}}\exp{-\frac{ab \e^{-2\im \alpha}}{1-b \e^{-2\im \alpha}}},\\
\sum_{n=0}^\infty \frac{1}{n+1}L_n^1(a)(be^{-2\im \alpha})^{n+1}=\frac{1}{a}\left[1-\exp{-\frac{abe^{-2\im \alpha}}{(1-be^{-2\im \alpha})}}\right],
    \end{align}
\end{subequations}
valid for $|b|<1$ as is the case here.
As a consequence, the unscattered amplitude \eqref{eq:psi0_in_defocus_plane_general} for any defocus angle and the scattering term $\Xi_z$ in \eqref{eq:scattpsi0_in_defocus_plane} become,
\begin{align}\label{eq:psi0_in_defocus_plane}
\la \vxi_z|\psi(0)\ra &=
\e^{-\im \alpha(z)}\sqrt{\frac{2}{\pi}}\frac{\kappa^2}{\kappa^{'2}\Delta_\e} \e^{-\kappa^2 r^2/2}\sum_{n=0}^\infty L_n(\kappa^2r^2)  \left(1-\frac{2\kappa^2}{\kappa^{'2}}\right)^n \e^{-2\im \alpha(z) n}\nonumber \\
&=\e^{-\im \alpha(z)}\sqrt{\frac{2}{\pi}}\frac{\kappa^2}{\kappa^{'2}\Delta_\e}\e^{-\kappa^2 r^2/2}\left(1-\left(1-\frac{2\kappa^2}{\kappa^{'2}}\right)\e^{-2\im \alpha(z)}\right)^{-1} \nonumber\\
&\quad \times\exp{-\kappa^2 r^2 \left(1-\frac{2\kappa^2}{\kappa^{'2}}\right)\e^{-2\im \alpha(z)} \left(1-\left(1-\frac{2\kappa^2}{\kappa^{'2}}\right)\e^{-2\im \alpha(z)}\right)^{-1}}
\end{align}
and \begin{equation}\label{eq:XIscattpsi0_in_defocus_plane}
    \begin{aligned}
\Xi_z(r)
=&-\im\frac{\kappa^2 r}{\sqrt{2\pi}} \frac{ \Delta_\s }{\Delta_\e} \e^{-\kappa^2 r^2/2}\sum_{n=0}^\infty \frac{1}{n+1} L_n^{1}(\kappa^2r^2)\left[\left(1-\frac{2\kappa^2}{\bar{\kappa}^{2}}\right)^{n+1}-\left(1-\frac{2\kappa^2}{\kappa^{'2}}\right)^{n+1} \right]\e^{-2\im \alpha(z)(n+1)}\\
=&-\frac{\im}{\sqrt{2\pi}\Delta_\e} \frac{ \Delta_\s }{ r} \e^{-\kappa^2 r^2/2}\left[\exp{-\kappa^2 r^2 \left(1-\frac{2\kappa^2}{\kappa^{'2}}\right)\e^{-2\im \alpha(z)} \left(1-\left(1-\frac{2\kappa^2}{\kappa^{'2}}\right)\e^{-2\im \alpha(z)}\right)^{-1}}\right.\\
&\qquad\qquad\qquad\qquad\qquad\left.-\exp{-\kappa^2 r^2 \left(1-\frac{2\kappa^2}{\bar{\kappa}^{2}}\right)\e^{-2\im \alpha(z)} \left(1-\left(1-\frac{2\kappa^2}{\bar{\kappa}^{2}}\right)\e^{-2\im \alpha(z)}\right)^{-1}}\right]
    \end{aligned}
\end{equation}
Notice that $\kappa$ acts like a scaling factor that can be absorbed in the position variables. 

\subsection{Classical trace distance for any defocus plane and diffraction-mode measurements}

Upon measuring in an arbitrary basis $\{ \ket{\vxi_z} \}$, the difference in outcome probabilities between perturbed and unperturbed states is governed by the overlap between $\la \vxi_z \vert \psi(0) \ra$ and $\la \vxi_z \vert \partial \psi(0) \ra$. Specifically, we find:
\begin{equation}
    \begin{split}
        P_{\rm NB}^{(B)}(\vxi_z)- P_{\rm NB}^{(A)}(\vxi_z)&= \left| \langle \vxi_z | \oS_{\rm NB}(\vartheta) | \psi(0) \rangle \right|^2 - \left| \langle \vxi_z | \psi(0) \rangle \right|^2 = 2\vartheta \Re\left\{ \langle \psi(0) | \vxi_z \rangle \langle \vxi_z |  \psi^\prime(0) \rangle \right\} + \mathcal{O}(\vartheta^2) \\ 
        &= 2\vartheta |\vn_\perp| \sin(\varphi)~\Re\left\{ \langle \psi(0) | \vxi_z \rangle \, \Xi_z(r) \right\} + \mathcal{O}(\vartheta^2).
    \end{split}
\end{equation}
The corresponding classical trace distance, according to Eq.~\eqref{eq:Basics:ClTrDist}, is:
\begin{equation}\label{eq:App:Disc:trDist_classical}
    D_{\rm NB}^{(z)} = 4\vartheta |\vn_\perp| \int r \, \mathrm{d}r \, \left|\Re\left\{ \langle \psi(0) | \vxi_z \rangle \, \Xi_z(r) \right\}\right| + \mathcal{O}(\vartheta^2),
\end{equation}
where we use radial symmetry, i.e., $\langle \vxi_z | \psi(0) \rangle$ depends only on $r$ for a cylindrically symmetric incident wavefunction.

Let us now examine the two special cases corresponding to $z = 0$ ($\alpha=0$, position measurement, image mode) and $z = f$ ($\alpha=\pi/2$, momentum measurement, diffraction mode). In the former case, we can immediately conclude that $ D_{\rm NB}^{(z=0)} = D_{\rm NB}^{\rm (x)} = 0$, because the scattering operator merely imprints a position-dependent phase which disappears when taking the absolute square of the wavefunction for the likelihood of outcomes. In Eq.~\eqref{eq:App:Disc:trDist_classical}, one sees this by noticing that $\langle \vxi_0 | \psi(0) \rangle \in \mathbb{R}$ and $\Xi_0(r) \in \im\mathbb{R}$.

In diffraction mode ($z = f$), we can infer the radial dependence of the momentum amplitude from \eqref{eq:scattpsi0_in_defocus_plane}, which holds for radial spin distributions and radial incident wavefunctions. Given $\vp_{\perp} = \hbar \kappa^2 \vxi_f$ in polar coordinates $(p,\varphi)$, 
\begin{equation}
    \begin{aligned}
\Xi_f(p)=\frac{\la \vp_\perp|\partial \psi(0)\ra}{|\vn_\perp| \sin(\varphi_p)}=-\frac{\im}{\sin(\varphi_p)}\int_0^\infty \diff r ~r \int_0^{2\pi} \diff \varphi~\frac{ \e^{\frac{\im}{\hbar}\vp_\perp\cdot\vx_\perp}}{2\pi \hbar} g(r) \sin(\varphi) \psi_0(r)
=\frac{1}{\hbar} \int_0^\infty \diff r~r J_1\left(\frac{p r}{\hbar}\right) g(r) \psi_0(r),
    \end{aligned}
\end{equation}
with $\psi_0(r)=\bra{\vx_\perp}\psi(0)\ra$. 
Consequently, $\langle \psi(0) | \vp_\perp \rangle \, \Xi_f(p)\in \mathbb{R}$ and 
\begin{equation}\label{eq:App:Disc:NBmom_general}
    D_{\rm NB}^{(z=f)} = D_{\rm NB}^{\rm (p)} =\vartheta |\vn_\perp|\cD  + \mathcal{O}(\vartheta^2), \quad \text{where} \quad
    \cD= 4\int _0^\infty \diff p ~p  \left|\langle \psi(0) | \vp_\perp \rangle \, \Xi_f(p)\right|.
\end{equation}
For the Gaussian model, we insert \eqref{eq:psi0_in_defocus_plane} to obtain:
\begin{subequations}\label{eq:App:DiscNB:MomAmplitudes}
    \begin{align}
        \langle \vxi_f | \psi(0) \rangle &= -\im \hbar \kappa^2 \bra{\vp_\perp}\psi(0)\ra = -\im  \kappa^2 \Delta_\e \sqrt{\frac{2}{\pi}}  \exp\left(-\frac{p^2 \Delta_\e^2}{\hbar^2} \right), \\
        \Xi_f(r) &= -\im \hbar \kappa^2 \Xi_f(p) = -\frac{\im \hbar \kappa^2\Delta_\s}{\sqrt{2\pi} \Delta_\e p} \left[ \exp\left(-\frac{p^2 \tilde{\Delta}_\e^2}{\hbar^2} \right) - \exp\left(-\frac{p^2 \Delta_\e^2}{\hbar^2} \right) \right],
    \end{align}
\end{subequations}
where the effective width is defined as $\tilde{\Delta}_\e^{-2} = \Delta_\e^{-2} + 2\Delta_\s^{-2}$. 
Then, 
\begin{equation}\label{eq:App:Disc:NBmom}
        \cD = \frac{4\Delta_\s}{\pi\hbar} \int _0^\infty \mathrm{d}p \, \exp\left(-\frac{p^2 \Delta_\e^2}{\hbar^2} \right)  \left[ \exp\left(-\frac{p^2 \tilde{\Delta}_\e^2}{\hbar^2} \right) - \exp\left(-\frac{p^2 \Delta_\e^2}{\hbar^2} \right) \right]  
        =  \sqrt{\frac{2}{\pi}} \frac{1}{\chi} \left( \sqrt{\frac{1 + 2\chi^2}{1 + \chi^2}} - 1 \right).
\end{equation}

\section{Discrimination with backaction}

\subsection{Quantum trace distance in the weak-sample limit}\label{App:QTrDist}

Our first goal is to obtain a perturbative expression for the quantum trace distance in the case of a magnetic spin with backaction. The reduced scattered electron state is no longer pure, and trace distances between mixed infinite-dimensional states often do not reduce to a simple expression. Here however, we can restrict to an effectively four-dimensional subspace spanned by the unscattered state and the OAM components of the scattered state. Given a general magnetic spin state with Bloch vector $\vc$ and a pure and radially symmetric incident electron state, $\orho_\e(0)=\ket{\psi(0)}\bra{\psi(0)}$, the scattered state \eqref{eq:finalstate_BA} can be expanded as:
 \begin{equation}\label{eq:rhoBA_expand}
 \hat{\rho}_\e (\vartheta)=\ket{0}\bra{0}+\frac{1+c_z}{2}\ket{+}\bra{+}+\frac{1-c_z}{2}\ket{-}\bra{-} + \left[ \frac{c_x+\im c_y}{2}\left(\ket{-}\bra{0}-\ket{0}\bra{+}\right) + h.c.\right],
 \end{equation}
 with the unnormalized vectors $\ket{0}=\hat{K}_0\ket{\psi(0)}$ and $\ket{\pm}=\hat{K}_\pm\ket{\psi(0)}$. The latter two are OAM eigenstates with eigenvalues $\pm \hbar$, and they are thus mutually orthogonal and orthogonal to $\ket{0}$ and $\ket{\psi(0)}$. We can normalize them as
 \begin{equation}
     |\psi_{\pm}\ra = \frac{|\pm\ra}{\sqrt{\la \psi(0)|\oK_{\pm}^\da \oK_{\pm}|\psi(0)\ra } } = \frac{\sin[\vartheta g(\opr)]e^{\pm\im \ovarphi}|\psi(0)\ra}{ \sqrt{\la \psi(0)| \sin^2[\vartheta g(\opr)] |\psi(0)\ra } } = \vartheta \frac{g(\opr)e^{\pm\im\ovarphi}|\psi(0)\ra}{\sqrt{\cG/2}} + \Order(\vartheta^3).
 \end{equation}
 The vector $|0\ra = \cos[\vartheta g(\opr)]|\psi(0)\ra$ can be expanded into the unscattered state and its orthogonal complement. For small $\vartheta$, 
\begin{equation}
    \begin{split}
\ket{0}\approx\ket{\psi(0)} -\frac{1}{2}\vartheta^2g^2(\opr)\ket{\psi(0)}+\cO(\vartheta^4)=\left(1-\frac{1}{4}\vartheta^2\cG\right)\ket{\psi(0)}-\frac{1}{2}\vartheta^2\bra{\psi_2}g^2(\hat{r})\ket{\psi (0)} \ket{\psi_2}+\cO(\vartheta^4)
    \end{split}
\end{equation}
where $\ket{\psi_2}$ denotes the normalized zero-OAM state orthogonal to $\ket{\psi (0)}$. Explicitly, 
\begin{equation}
    \ket{\psi_2}=\frac{g^2(\opr)-(\cG/2)^2}{ \sqrt{\expval{g^4(\opr)}_0-(\cG/2)^2 }}\ket{\psi(0)}.
\end{equation}
Introducing $\bar{\vartheta}=\vartheta\sqrt{\cG/2}$, we then have the following approximation to second order:
\begin{equation}\label{eq:vvar_general}
    \ket{0} =\left(1-\frac{\bar{\vartheta}^2}{2}\right)\ket{\psi(0)}- \cV \bar{\vartheta}^2\ket{\psi_2}+\cO(\vartheta^4), \qquad \text{with} \quad \cV = \frac{1}{2}\sqrt{\frac{\cF^2}{\cG^2}-1}\quad \text{and}\quad \cF=2\sqrt{\expval{g^4(\opr)}_0}
\end{equation}
This allows us to express the scattered density matrix \eqref{eq:rhoBA_expand} to second order in $\vartheta$ in the basis $\{\ket{\psi(0)}, \ket{\psi_2}, \ket{\psi_+}, \ket{\psi_-}\}$, 
\begin{equation}\label{eq:rhoBA:secondorder}
    \orho_\e(\vartheta)=\begin{pmatrix}
        1-\bar{\vartheta}^2 & -\bar{\vartheta}^2\cV & \frac{|\vc_\perp|}{2} \e ^{\im \varphi_c}\bar{\vartheta} & \frac{|\vc_\perp|}{2} \e ^{-\im \varphi_c}\bar{\vartheta} \\
        -\bar{\vartheta}^2\cV & 0 & 0 & 0\\
        \frac{|\vc_\perp|}{2} \e ^{-\im \varphi_c}\bar{\vartheta} & 0 & \frac{1}{2}(1+c_z)\bar{\vartheta}^2 & 0\\
        \frac{|\vc_\perp|}{2} \e ^{\im \varphi_c}\bar{\vartheta} & 0 & 0 & \frac{1}{2}(1-c_z)\bar{\vartheta}^2
    \end{pmatrix}+\cO(\vartheta^3),
\end{equation}
where $\vc_{\perp} = |\vc_{\perp}| (\cos\varphi_c \ve_x + \sin\varphi_c \ve_y)$. The dependence on the polar angle $\varphi_c$ of the spin Bloch can be removed by applying a unitary transformation, 
\begin{equation}\label{eq:app:unitary_rotation}
    \oU=\begin{pmatrix}
        1 & 0 & 0 & 0 \\
        0 & 1 & 0 & 0 \\
        0 & 0 & \frac{1}{\sqrt{2}}\e^{\im \varphi_c} & \frac{1}{\sqrt{2}}\e^{-\im \varphi_c}\\
        0 & 0 & \frac{1}{\sqrt{2}}\e^{\im \varphi_c} & -\frac{1}{\sqrt{2}}\e^{-\im \varphi_c}
    \end{pmatrix}, \qquad \text{so that}\quad \oU\orho_\e(\vartheta)\oU^\dagger=\begin{pmatrix}
        1-\bar{\vartheta}^2 & -\bar{\vartheta}^2\cV & 0 & \frac{|\vc_\perp|}{\sqrt{2}}\bar{\vartheta} \\
        -\bar{\vartheta}^2\cV & 0 & 0 & 0\\
        0 & 0 & \frac{1}{2}\bar{\vartheta}^2 & \frac{1}{2}c_z\bar{\vartheta}^2\\
        \frac{|\vc_\perp|}{\sqrt{2}}\bar{\vartheta} & 0 & \frac{1}{2}c_z\bar{\vartheta}^2 & \frac{1}{2}\bar{\vartheta}^2
    \end{pmatrix}+\cO(\vartheta^3).
\end{equation}
This implies that the quantum trace distance and the QFI do not depend on the in-plane orientation of the Bloch vector either.

The term $\cV$ in \eqref{eq:vvar_general} depends on the expectation value $\expval{g^4(\opr)}_0$ and on $\cG = 2\expval{g^2(\opr)}_0$, which must be computed numerically in general. In the Gaussian model with standard deviations $\Delta_\e$ (electron wavefunction) and $\Delta_\s$ (spin size), there exists an analytic expression in terms of $\chi = \Delta_\e/\Delta_\s$, determined by $g(\opr)$ in \eqref{eq:g_Gauss}, $\cG$ in \eqref{eq:cG:Gaussian}, and by
\begin{equation}\label{eq:G4}
    \begin{aligned}
\expval{g^4(\opr)}_0=&\int_0^\infty \diff r \frac{\Delta_\s^4}{\Delta_\e^2 r^3} \left[1-\exp\left(-\frac{r^2}{2\Delta_\s^2}\right)\right]^4 \exp\left(-\frac{r^2}{2\Delta_\e^2}\right)\\
=& \frac{1}{4\chi^4}\left[\log\left(\frac{(1+2 \chi^2)^3}{(1+\chi^2)^3(1+3\chi^2)}\right)+(4\chi^2+1)\log\left(\frac{(4\chi^2+1)(2\chi^2+1)^3}{(\chi^2+1)(1+3\chi^2)^3}\right)\right].
    \end{aligned}
\end{equation}

\begin{figure}
    \centering
    \includegraphics[width=0.5\linewidth]{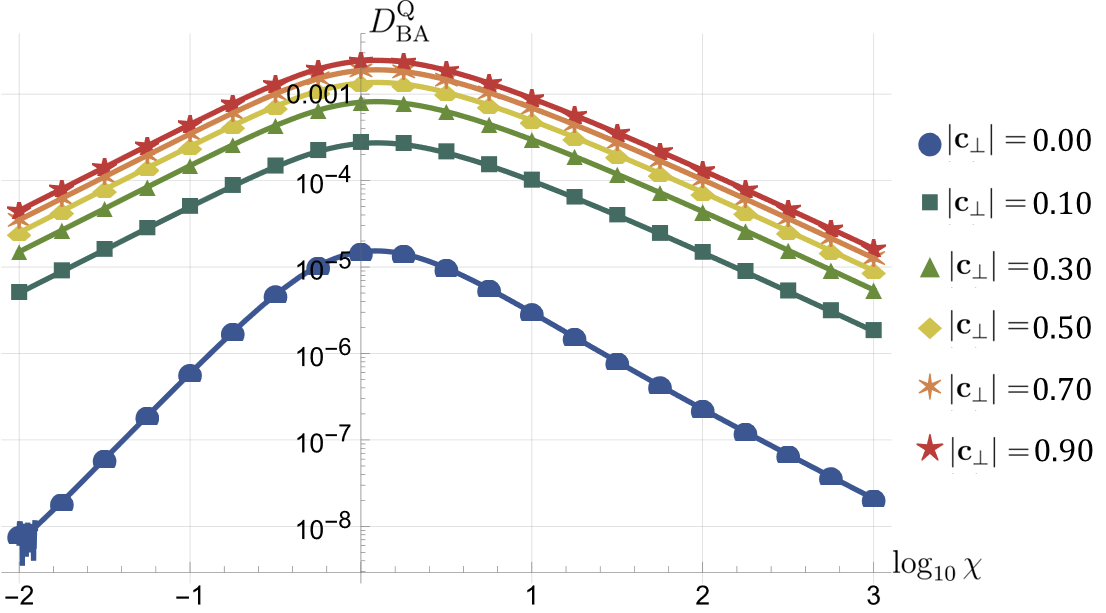}
    \caption{Quantum trace distance as a function of $\chi$ for $\vartheta=10^{-2}$ and fixed $c_z=0.1$, evaluated for different values of $|\vc_\perp|$ on the Bloch sphere. Dots represent the numerical results obtained from the spectral decomposition, while the continuous line corresponds to Eq.~\eqref{eq:App:DiscBA:QtrDist}, demonstrating agreement with the perturbative eigenvalue analysis.}
    \label{fig:DQ_BA}
\end{figure}

With this, we can now evaluate the quantum trace distance between $\orho_\e (0)$ and $\orho_\e (\vartheta)$. Making use of the unitary transformation in \eqref{eq:app:unitary_rotation}, we have
\begin{equation}\label{eq:UrhoBA:secondorder}
    \oU\left[\orho_\e(\vartheta)-\ket{\psi(0)}\bra{\psi(0)}\right]\oU^\dagger=
    \begin{pmatrix}
        -\bar{\vartheta}^2 & -\bar{\vartheta}^2\cV & 0 & \frac{|\vc_\perp|}{\sqrt{2}}\bar{\vartheta} \\
        -\bar{\vartheta}^2\cV & 0 & 0 & 0\\
        0 & 0 & \frac{1}{2}\bar{\vartheta}^2 & \frac{1}{2}c_z\bar{\vartheta}^2\\
        \frac{|\vc_\perp|}{\sqrt{2}}\bar{\vartheta} & 0 & \frac{1}{2}c_z\bar{\vartheta}^2 & \frac{1}{2}\bar{\vartheta}^2
    \end{pmatrix}.
\end{equation}
The quantum trace distance is given in terms of the four eigenvalues of this matrix, 
$D_{\rm BA}^{\rm Q} = [|\lambda_1|+|\lambda_2|+|\lambda_3|+|\lambda_4|]/2$.
For $\vc_\perp=0$, the eigenvalues are: $\lambda_{1,2}=\frac{\bar{\vartheta}^2}{2}\left(-1\pm \sqrt{1+4\cV^2}\right)+\cO(\vartheta^4)$ and $\lambda_{3,4} = \frac{1}{2}(1 \pm c_z) \bar{\vartheta}^2$.
The trace distance then reduces to
\begin{equation}\label{eq:DiscBA:qtraceC0}
D_{\rm BA}^{{\rm Q}~(\vc_\perp=0)}
=\frac{1}{4}\vartheta^2 \cG\left[1+\sqrt{1+4\cV^2}\right]+\cO(\vartheta^4).
\end{equation}
For arbitrary initial spin states, expanding the eigenvalues in powers of $\vartheta$ gives: $\lambda_1 = \frac{1}{2}\bar{\vartheta}^2+\cO(\vartheta^3)$, $\lambda_2 = \frac{1}{\sqrt{2}}|\vc_\perp|\bar{\vartheta}-\frac{1}{4}\bar{\vartheta}^2+\cO(\vartheta^3)$, 
$\lambda_3 = -\frac{1}{\sqrt{2}}|\vc_\perp|\bar{\vartheta}-\frac{1}{4}\bar{\vartheta}^2+\cO(\vartheta^3),$ and $\lambda_4 = \cO(\vartheta^3)$.
To leading order, these eigenvalues are independent of $c_z$, implying that any spin state with the same transverse Bloch vector component produces the same deviation from the incident electron state, regardless of its alignment with the beam or bias field. The parameter $\cV$ only appears at order $\vartheta^3$ when $|\vc_\perp|\neq 0$.
An approximate expression for the quantum trace distance is therefore
\begin{equation}\label{eq:App:DiscBA:QtrDist}
D^{\rm Q}_{\rm BA}=\begin{cases}
    \frac{1}{2}\vartheta|\vc_\perp|\sqrt{\cG} +\cO(\vartheta^2), & \text{if } |\vc_\perp|>\frac{1}{\sqrt{8}}\vartheta,\\
    \frac{1}{4}\vartheta^2 \cG\left[1+\sqrt{1+4\cV^2}\right]+\cO(\vartheta^4), & \text{if } |\vc_\perp|=0.
\end{cases}
\end{equation}
For sufficiently small $\vartheta$, as expected for single-spin magnetic samples, the first case applies for almost any $|\vc_\perp|\neq 0$. Numerical evaluations confirm excellent agreement with Eq.~\eqref{eq:App:DiscBA:QtrDist} (see Fig.~\ref{fig:DQ_BA}). Unlike the no-backaction case, here the trace distance remains nonzero even for spins aligned with the beam propagation or bias field axis, highlighting the role of second-order terms in generating electron state decoherence.

\subsection{Classical trace distance with position and momentum measurements} 
\label{App:DiscBA:PosAndMom}

We now evaluate the classical trace distance between the probability distributions obtained from a position and momentum measurement of the unscattered state $\ket{\psi(0)}\bra{\psi(0)}$ and the scattered state $\orho_\e(\vartheta)$ in Eq~\eqref{eq:App:ScattChanSimplified}. 
In Eqs.~\eqref{eq:App:BA:ProbPos} and \eqref{eq:App:BA:ProbMom}, we derived explicit expressions for the probability density in position and momentum space, respectively, in the backaction scenario.  
From the position-space result, it follows directly that $P(\vx_\perp |\vartheta)
= P(\vx_\perp | 0)$, 
implying that the classical trace distance for a position measurement vanishes.  
In contrast, the momentum-space probability density can be treated perturbatively by expanding $\oK_0$ and $\oK_\pm$ to first order in~$\vartheta$.  
For the non scattered probability amplitude we have to first order
\begin{equation}
    \begin{aligned}
        \bra{\vp_\perp } \oK_0 | \psi(0)\ra
        &= \la{\vp_\perp | \psi(0)}\ra + \mathcal{O}(\vartheta^2) = \frac{\im}{\hbar \kappa^2} \, \la{\vxi_f | \psi(0)}\ra + \mathcal{O}(\vartheta^2) 
    \end{aligned}
\end{equation}
since $\oK_0 \approx \mathds{1}$ at this order.  
The overlap between the scattered component of the electron state and a transverse-momentum plane wave reads
\begin{equation}
    \begin{aligned}
        \la{\vp_\perp | \sin(\vartheta g(\opr)) \sin(\ovarphi) | \psi(0)}\ra
        &= \la{\vp_\perp | \vartheta g(\opr) \sin(\ovarphi) | \psi(0)}\ra + \mathcal{O}(\vartheta^3) = \frac{\im \vartheta}{|\vn_\perp|} \frac{\im}{\hbar \kappa^2} \, \la{\vxi_f | \partial \psi(0)}\ra + \mathcal{O}(\vartheta^3) \\
        &= -\frac{\vartheta}{\hbar \kappa^2} \sin(\varphi_p) \, \Xi_f(r) + \mathcal{O}(\vartheta^3) = \im \vartheta \sin(\varphi_p) 
        \Xi_f(p)
        + \mathcal{O}(\vartheta^3),
    \end{aligned}
\end{equation}
where $\varphi_p$ is the polar angle of the transverse momentum.  
Since also $| \la{\vp_\perp | \oK_{\pm} | \psi(0)}|^2 = \Order(\vartheta^2)$, Eq.~\eqref{eq:App:BA:ProbMom} becomes
\begin{equation}\label{eq:App:PBAMom:firstorder}
    \begin{split}
        P_{\mathrm{BA}}(\vp_\perp \mid \vartheta) 
        &= \left| \la{\vp_\perp}|{\psi(0)}\ra \right|^2 
        + 2 |\vc_\perp| \, \Im \left\{ 
        \la{\vp_\perp | \sin(\vartheta g(\opr)) \sin(\ovarphi) | \psi(0)}\ra
        \la{\psi(0)}|{\vp_\perp}\ra \right\} 
        + \mathcal{O}(\vartheta^2) \\
        &= \left| \la{\vp_\perp}|{\psi(0)}\ra \right|^2 
        + 2 \vartheta |\vc_\perp| \, \Re \left\{ 
        \Xi_f(p) \la{\psi(0)}|{\vp_\perp}\ra \right\} \sin(\varphi_p)
        + \mathcal{O}(\vartheta^2).
    \end{split}
\end{equation}
Identifying $P^{(A)}_{\mathrm{BA}}(\vp_\perp) = P_{\mathrm{BA}}(\vp_\perp \mid \vartheta)$ and $
P^{(B)}_{\mathrm{BA}}(\vp_\perp) = \left| \braket{\vp_\perp}{\psi(0)} \right|^2$,
the trace distance to second order in~$\vartheta$ reads
\begin{equation}
    \begin{aligned}
        D^{(\mathrm{p})}_{\mathrm{BA}} 
        &= \vartheta |\vc_\perp| \int_0^\infty \! \mathrm{d}p \, p \int_0^{2\pi} \! \mathrm{d}\varphi_p \,
        |\sin(\varphi_p)| \, 
        \left| \Re\left\{ \Xi_f(p) \braket{\psi(0)}{\vp_\perp} \right\} \right|
        + \mathcal{O}(\vartheta^2) \\
        &= 4\vartheta |\vc_\perp| \int_0^\infty \! \mathrm{d}p \, p \, \Xi_f(p) \braket{\psi(0)}{\vp_\perp}
        + \mathcal{O}(\vartheta^2)= \vartheta |\vc_\perp| \, \mathcal{D} + \mathcal{O}(\vartheta^2)
    \end{aligned}
\end{equation}
in analogy to Eq.~\eqref{eq:App:Disc:NBmom_general}.
For an incident Gaussian electron beam and a Gaussian spin density, we recover $\cD$ from Eq.~\eqref{eq:App:Disc:NBmom}.
Consequently, the quantum and classical trace distances from a momentum measurement are unaffected by backaction effects to first order in~$\vartheta$.  
In both cases, the quantum limit is not saturated by a momentum measurement.  
In the backaction case, scanning over different defocus planes requires further analysis, since the energy exchange between the electron and the quantum spin creates a superposition of electron wavefunctions with different energies.  
The optical unitary used in the no-backaction scenario then acts differently on each energy component, introducing additional phase delays associated with the energy gain or loss.

\section{Testing the validity of the SVEA}\label{App:Comparison}

From a parallel work~\cite{simulationSRS_TEM}, we developed a full three-dimensional approach to the scattering process, including backaction effects. In that work, we applied the paraxial approximation only to the incident electron wavefunction, which is a valid assumption for $\Delta_\e > 10^{-12}~\mathrm{m}$ and a beam energy of 200~keV, since $\Delta_\e p_0/\hbar \gg 1$ with $p_0/\hbar \sim 10^{13}~\mathrm{m}^{-1}$. The range of validity of the SVEA is then tested by comparing the two approaches to the scattering problem.  

As the analysis in~\cite{simulationSRS_TEM} was restricted to the first-order contribution of the interaction between free electrons and a quantum spin, we can in particular compare the probability amplitudes of the scattered part of the final electron state. These amplitudes are then reflected in the probability distribution both in position and momentum space. Since the relevant contribution appears in momentum measurements of a spin transversally aligned with respect to the beam propagation axis, we focus on this case. Without loss of generality, we study a spin fully aligned along the transversal plane, i.e. $c_z=0$. Retrieving the probability from~\cite{simulationSRS_TEM} and adopting the notation of the present work, we obtain
\begin{equation}
    P_{\rm BA}^{(\rm 3D)}(\vp_\perp|\vartheta)=|\la \vp_\perp|\psi(0)\ra|^2+|\vc_\perp|\la \psi(0)|\vp_\perp\ra\Re\left\{{\beta_{\|, +}(\vp_\perp; 0, 0)+\beta_{\|, -}}(\vp_\perp; 0, 0)\right\}+\cO(\vartheta^2),
\end{equation}
where the coefficients $\beta_{\|, \pm}$ are given explicitly in Eq. (C.35) for $t=0$ in the precession evolution and evaluated in the far field.

This probability distribution can be directly compared with Eq.~\eqref{eq:App:PBAMom:firstorder}, where we note in addition that $\Xi_f(p)\in \mathbb{R}$ and $\la \vp_\perp|\psi(0)\ra\in \mathbb{R}$.
On the other hand, using the expressions $\ell(p; 0, 1)$ in Eq.~(C.31) for $z=0$, we have also that $\beta_{\|, +}(\vp_\perp; 0, 0)+\beta_{\|, -}(\vp_\perp; 0, 0) \in \mathbb{R}$. Taking the following from Eq.~(C.35) we found that
\begin{equation}
\beta_{\|, +}(\vp_\perp; 0, 0)+\beta_{\|, -}(\vp_\perp; 0, 0)=2\vartheta\left[\sqrt{\frac{2}{\pi}}\frac{\Delta_\s}{\hbar}\ell(p; 0, 1)\right]\sin(\varphi_p) = 2\vartheta \Xi_f^{(\rm 3D)} (p) \sin(\varphi_p).
\end{equation}
The square-bracketed term $\Xi_f^{(\rm 3D)}$ replaces $\Xi_f$ in \eqref{eq:App:PBAMom:firstorder}. We split it into parallel and perpendicular contributions,
\begin{equation}
    \Xi_f^{(\rm 3D)}(p)=\Xi_\perp(p,f) + \Xi_\|(p,f)=\sqrt{\frac{2}{\pi}}\frac{\Delta_\s}{\hbar}\ell_\perp(p; 0,1) + \sqrt{\frac{2}{\pi}}\frac{\Delta_\s}{\hbar}\ell_\|(p; 0,1),
\end{equation}
which reflects the orientation of the magnetic field relative to the beam electron. For large probe sizes, the perpendicular component dominates, since the spread in transverse momentum remains narrow. As the probe size decreases, however, the electron wavefunction acquires a broader transverse momentum distribution, making it increasingly sensitive to the longitudinal component of the magnetic field.

\begin{figure}
    \centering
    \includegraphics[width=0.75\linewidth]{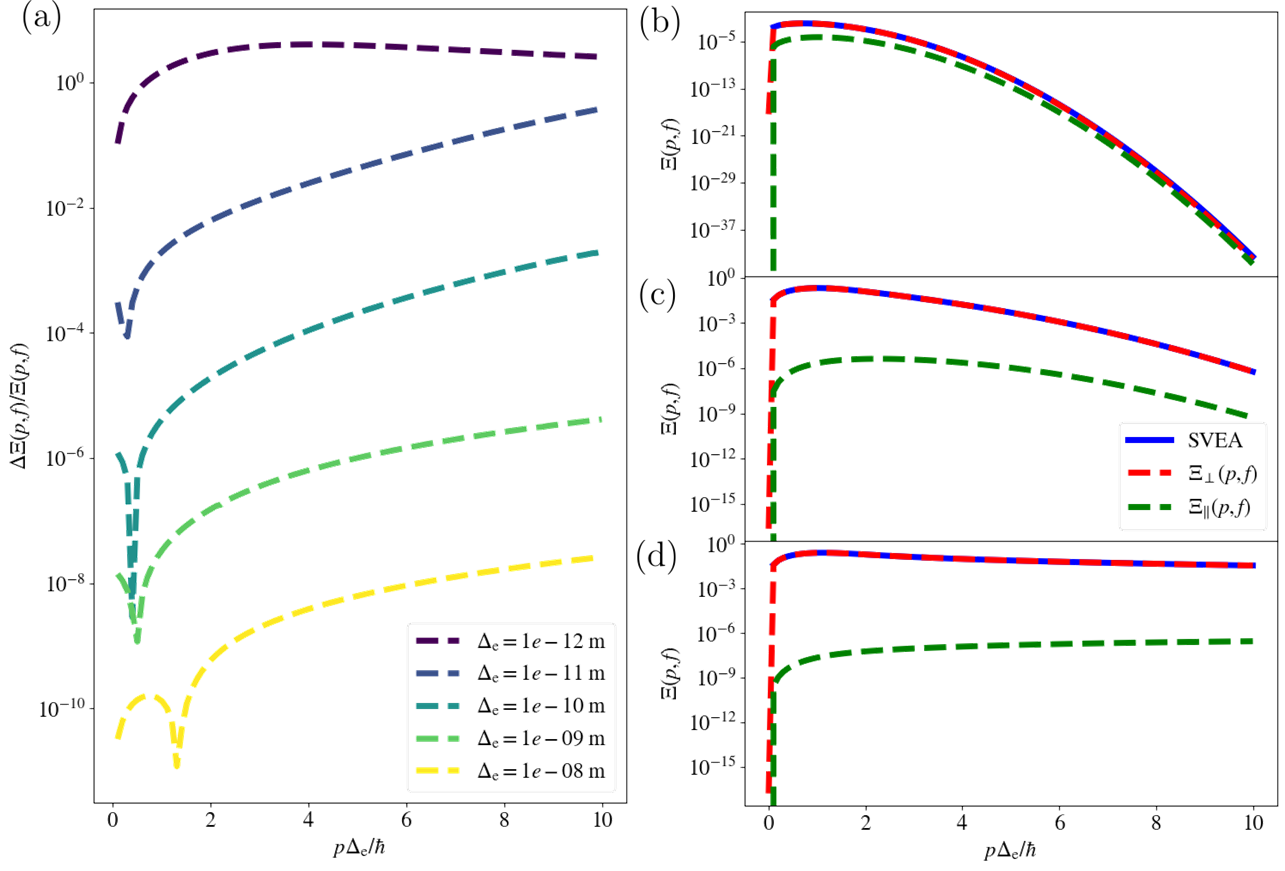}
    \caption{(a) Relative differences of $\Xi_f(p)$ from the SVEA with respect to the full radial amplitude $\Xi_f^{\rm (3D)}(p)$ are shown as a function of $p$ for different probe sizes, assuming a smeared spin of size $\Delta_\s=50$~pm. The radial components of the scattering amplitude in the SVEA (dashed blue), $\Xi_\perp(p,f)$ (dashed red), and $\Xi_\|(p,f)$ (dashed green) are displayed for (b) $\Delta_\e=10^{-11}$~m, (c) $\Delta_\e=10^{-10}$~m, and (d) $\Delta_\e=10^{-9}$~m. 
For very small probe sizes, near the limit of the paraxial approximation, the longitudinal contribution becomes only about one order of magnitude smaller than the transverse part, so the SVEA ceases to provide a reliable description of the scattering process. In contrast, the transverse component shows consistent overlap for different $\Delta_\s$, confirming that the SVEA accurately captures this contribution. At large $p$, the relative difference decreases, but the radial amplitudes are exponentially suppressed, limiting the amount of extractable information in that regime. }
    \label{fig:comparison_SVEA_3D}
\end{figure}

Implementing the numerical code to evaluate the integrals $\ell_\|(p;0,1)$ and $\ell_\perp(p;0,1)$ (Eq.~(C.31)), we obtain the results shown in Fig.~\ref{fig:comparison_SVEA_3D}. In Fig.~\ref{fig:comparison_SVEA_3D}a we compare the radial part of the scattering amplitude in momentum space to first order for different probe and smearing sizes, through the relative deviation
\[
\frac{\Delta\Xi_f(p)}{\Xi_f(p)}=\frac{|\Xi_f^{(\rm 3D)}(p)-\Xi_f(p)|}{|\Xi_f^{(3D)}(p)|+|\Xi_f(p)|}.
\]
We find that the amplitudes differ by roughly one order of magnitude for any $\Delta_\s$, or less, once the probe size approaches the paraxial limit, $\Delta_\e \sim 10^{-12}$~m. This discrepancy becomes especially relevant when quantifying the information content in diffraction-mode imaging, since these techniques rely on the full momentum distribution. For a spin with a smearing size of $50$~pm, we therefore restrict the plot to $\chi > 10^{-1}$. As shown in Fig.~\ref{fig:comparison_SVEA_3D}b--d, a probe size $\Delta_\e\sim1$~pm leads to a significant contribution from the parallel components, which eventually dominates over the transverse part. In general, the SVEA overlaps with the perpendicular contribution, as this is precisely the term retained in the paraxial approximation. Small deviations appear at very small probe sizes, where energy transfer to non-negligible transverse components of the beam becomes important. 

If we approximate $\ell_\perp(p;0,1)$ using Eq.~(C.45), under paraxial conditions on the electron state, the relevant integral for a Gaussian smearing function in terms of dimensionless variables $\tilde{q}=\Delta_\s q/\hbar$ and $\tilde{p}=\Delta_\s p/\hbar$ reads
\begin{equation}
    \begin{aligned}
\ell_\perp(p, 0)
=&\chi \int_0^\infty \diff \tilde{q}~ \exp\left(-\chi^2(\tilde{p}^2+\tilde{q}^2)\right)  \exp\left(-\tfrac{1}{2}\tilde{q}^2\right)I_1\!\left(2\chi^2\tilde{p}\tilde{q}\right)\\
=&\chi \exp\!\left(-\chi^2\tilde{p}^2\right)\frac{1}{2\chi^2 \tilde{p}}\left[-1+\exp\!\left(\tfrac{2 \chi^4\tilde{p}^2}{1+2\chi^2}\right)\right]=\frac{\hbar}{2\Delta_\e p}\left[\exp\!\left(-\tfrac{\chi^2}{1+2\chi^2}\tilde{p}^2\right)-\exp\!\left(-\chi^2\tilde{p}^2\right)\right].
    \end{aligned}
\end{equation}
From the last expression it follows directly that
\begin{equation}
    \Xi^{\rm (3D)}_\perp=\frac{1}{\sqrt{2\pi}\chi p}\left[\exp\!\left(-\frac{\tilde{\Delta}_\e^2p^2}{\hbar^2}\right)-\exp\!\left(-\frac{\Delta_\e^2p^2}{\hbar^2}\right)\right]=\Xi_f(p),
\end{equation}
with $\tilde{\Delta}_\e^2=\chi^2\Delta_\s^2/(1+2\chi^2)$. Thus, assuming that the electron state has a narrow longitudinal momentum spread and transverse components much smaller than the longitudinal one (valid for $\Delta_\e>10^{-12}$~m), we recover the SVEA result in momentum space. The main corrections arise from $\ell_\|(p;0,1)$, since under the SVEA one obtains $f_\|(p,0)=0$, highlighting that the parallel part accounts for the discrepancy relative to the full momentum-space treatment.  

Near $\Delta_\e \sim 10^{-12}$~m, further corrections become relevant. In Ref.~\cite{simulationSRS_TEM}, the longitudinal momentum of the incident beam is assumed to be sufficiently narrow to approximate it by its central value-an assumption that breaks down for extremely focused beams. Moreover, the expansion in small momentum must then retain the full expressions, as shown in Eq.~(C.26) of \cite{simulationSRS_TEM}.  
For imaging other samples and their propagation to different planes, the same reasoning applies by including the defocus propagation, where the beam acquires an average phase delay after interaction. This phase encodes the energy exchanged between quantum systems as well as between the momentum degrees of freedom of the electron beam, with the averaging arising from coherent superposition of different plane-wave components.

\section{Examples of Spin Sensing in Magnetic Dipole Samples}\label{App:Real}

As shown previously, the quantum trace distance and the QFI depend on the variable $\cG$, which quantifies the averaged interaction with respect to the spin density, smeared out in position, and the spatial shape of the incident wavefunction, assuming that both are radially symmetric. In general, from Eqs.~\eqref{eq:App:general_g_func} and \eqref{eq:QFI:NB}, we can write
\begin{equation}
\begin{aligned}
\cG &= \frac{2\pi^2}{\chi^2} \int_0^\infty \diff r \, r^3 \exp\left(-\frac{r^2}{2\Delta_\e^2}\right) 
\int_{-\infty}^{+\infty} \frac{\diff z}{\sqrt{r^2+z^2}}
\int_0^\infty \diff u \int_{-1}^1 \diff \tau \, \tau P_\s(\scriptr(r, z, u, \tau)) \\
&\quad \times \int_{-\infty}^{+\infty} \frac{\diff z'}{\sqrt{r^2+z'^2}}
\int_0^\infty \diff u' \int_{-1}^1 \diff \tau' \, \tau' P_\s(\scriptr(r, z', u', \tau)),
\end{aligned}
\end{equation}
where $\scriptr(r, z, u, \tau) = \sqrt{r^2 + z^2 + u^2 + 2 \sqrt{r^2 + z^2} \, u \, \tau}$.

\paragraph{Magnetic ball}\label{App:Real:Sphere}

For a magnetic moment distributed over a solid sphere of radius $R$, the spin density is given by $P_\s(r) = (3/4\pi R^3) \Theta(R-r)$,
with $\Theta$ the Heaviside function. 
Introducing dimensionless position coordinates in units of $R$, $r = \tilde{r} R,$ $u = \tilde{u} R,$ $ z = \tilde{z} R,$ $ u' = \tilde{u}' R,$ $ z' = \tilde{z}' R$, we obtain
\begin{equation}
\begin{aligned}
\cG &= \frac{9}{8\chi^2} \int_0^1 \diff \tilde{r} \, \tilde{r}^3 \exp\left(-\frac{\tilde{r}^2}{2\chi^2}\right) 
\int_{-1}^{1} \frac{\diff \tilde{z}}{\sqrt{\tilde{r}^2 + \tilde{z}^2}} 
\int_0^1 \diff \tilde{u} \int_{-1}^1 \diff \tau \, \tau \, \Theta(1 - \scriptr(\tilde{r}, \tilde{z}, \tilde{u}, \tau)) \\
&\quad \times \int_{-1}^{1} \frac{\diff \tilde{z}'}{\sqrt{\tilde{r}^2 + \tilde{z}'^2}} 
\int_0^1 \diff \tilde{u}' \int_{-1}^1 \diff \tau' \, \tau' \, \Theta(1 - \scriptr(\tilde{r}, \tilde{z}', \tilde{u}', \tau)).
\end{aligned}
\end{equation}
Here, we can use finite integration limits, because the integrals contribute negligibly when $r, |z|, u > R$ and thus $\scriptr(\tilde{r}, \tilde{z}, \tilde{u}, \tau) > 1$ for all $\tau$. Monte Carlo integration of $\cG$, yields the results shown in Figure~\ref{fig:realistic_NeNshots}(a).

\paragraph{Orbital electron}\label{App:Real:H1s}
Next, we consider an electron spin delocalized over the 1s orbital of the hydrogen atom, $P_\s (r) \exp{-2r/a_0}/\pi a_0^3$, with $a_0$ the Bohr radius. In this case, we obtain an analytic expression for $g(r)$ as follows. First, we define $\tilde{\rho}=\sqrt{\opr^2/a_0^2+\tilde{z}^2}$ and write,
\begin{equation}
\begin{aligned}
g(r ) =& -\pi a_0 r  \int_{-\infty}^{+\infty}\frac{\diff z}{\sqrt{r^2+z^2}}\int_0^\infty \diff u \int_{-1}^1 \diff \tau ~\tau \frac{1}{\pi a_0^3}\exp\left(-\frac{2}{a_0}\sqrt{r^2+u^2+z^2+2u\sqrt{r^2+z^2}\tau}\right)\\
=& -\frac{r}{a_0} \int_{-\infty}^{+\infty}\frac{\diff \tilde{z}}{\tilde{\rho}}\int_0^\infty \diff \tilde{u} \int_{-1}^1 \diff \tau ~\tau \exp\left(-2\sqrt{\tilde{\rho}^2+\tilde{u}^2+2\tilde{u}\tilde{\rho}\tau}\right) .
\end{aligned}
\end{equation}
A change of variable from $\tau$ to $s=\sqrt{\tilde{\rho}^2+\tilde{u}^2+2\tilde{\rho}\tilde{u} \tau}$, with $s\in [|\tilde{\rho}-\tilde{u}|,\tilde{\rho}+\tilde{u}]$ and $s \diff s=\tilde{\rho}\tilde{u}\diff \tau$, leads to
\begin{equation}
g(r) = \frac{r}{a_0} \int_{-\infty}^{+\infty}\frac{\diff \tilde{z}}{\tilde{\rho}}\int_0^{\infty} \diff \tilde{u}\int_{|\tilde{u}-\tilde{\rho}|}^{\tilde{u}+\tilde{\rho}} \diff s \frac{s(\tilde{\rho}^2+\tilde{u}^2-s^2)}{2\tilde{\rho}^2\tilde{u}^2}\e^{-2s} 
= \frac{r}{a_0} \int_{-\infty}^{+\infty}\frac{\diff \tilde{z}}{2\tilde{\rho}^3} \int_0^\infty \diff s ~s\e^{-2 s}\left.\left[\tilde{u}-\frac{1}{\tilde{u}}(\tilde{\rho}^2-s^2)\right]\right|^{s+\tilde{\rho}}_{|s-\tilde{\rho}|}.
\end{equation}
Here, since the $(\tilde{u},s)$-integration is over the region $\{(\tilde{u},s):\tilde{u}\geq0, s\geq0, |\tilde{\rho}-\tilde{u}|\leq s \leq \tilde{\rho}+\tilde{u}\} = \{(\tilde{u},s):\tilde{u}\geq0, s\geq0, |\tilde{\rho}-s|\leq \tilde{u} \leq \tilde{\rho}+s\}$, we could switch the integration order and carry out the $\tilde{u}$-integral. The resulting function of $\tilde{u}$  evaluates to $4s$ if $s \leq \tilde{\rho}$ and to zero otherwise, so that  
\begin{align}
g(r) &= \frac{2r}{a_0}\int_{-\infty}^{+\infty}\frac{\diff \tilde{z}}{\tilde{\rho}^3} \int_0^{\tilde{\rho}} \diff s~ s^2 \e^{-2s}= \frac{r}{2 a_0}\int_{-\infty}^{+\infty}\frac{\diff \tilde{z}}{\tilde{\rho}^3}\left[1-(2\tilde{\rho}^2 + 2\tilde{\rho} +1 )\e^{-2\tilde{\rho}}\right] \nonumber \\
&= \frac{a_0}{r} - \int_{-\infty}^{+\infty} \diff\theta \left[ \frac{r}{a_0} +\sech(\theta)+\frac{a_0}{2r} \sech^2 (\theta) \right]\exp\left[-\frac{2r}{a_0} \cosh(\theta)\right] .
\end{align}
In the second line we have evaluated the first term of the $\tilde{z}$-integral in the first line, and for the remaining terms, we have substituted $\tilde{z} (\theta)=r\sinh(\theta)/a_0$. 
We now introduce the integral representation of modified Bessel functions $K_{\nu}$, define an auxiliary integral function $J$, and notice an integral identity,
\begin{equation} 
K_\nu(b)=\int_0^{\infty} \!\!\!\diff x \, \cosh(\nu x) \, \e^{-b \cosh x}, \quad  J(b)=\int_0^{\infty} \!\!\! \diff x \, \sech(x) \, \e^{-b \cosh x }, \quad \int_0^{\infty}\!\!\! \diff x \, \sech^2(x) \e^{-b \cosh x} = b [K_1(b) + J(b)].
\end{equation}
This allows us to obtain the final result,
\begin{align}
 g(r) = \frac{a_0}{r} - \frac{2r}{a_0} K_0\left(\frac{2r}{a_0}\right) - 2 K_1\left(\frac{2r}{a_0}\right),   
\end{align}
which scales like $g(r) \rightarrow (r/a_0)\ln(r/a_0)$ for small $r$, 
removing the divergence at $r=0$ for a localized spin.
Setting $\chi=\Delta_\e/a_0$, we get 
\begin{equation}
\cG = \frac{1}{\chi^2} \int_0^{\infty} \diff \tilde{r}~\tilde{r} \exp\left(-\frac{r^2}{2\chi^2}\right)\left(\frac{1}{\tilde{r}}-2\tilde{r}K_0\left(2\tilde{r}\right)-2K_1\left(2\tilde{r}\right)\right)^2.
\end{equation}
The numerical solution of this integral is shown in Figure~\ref{fig:realistic_NeNshots}(b).

\paragraph{Nuclear spin}\label{App:Ne:NuclearSpins}

\begin{figure}[h]
    \centering
    \includegraphics[width=0.75\linewidth]{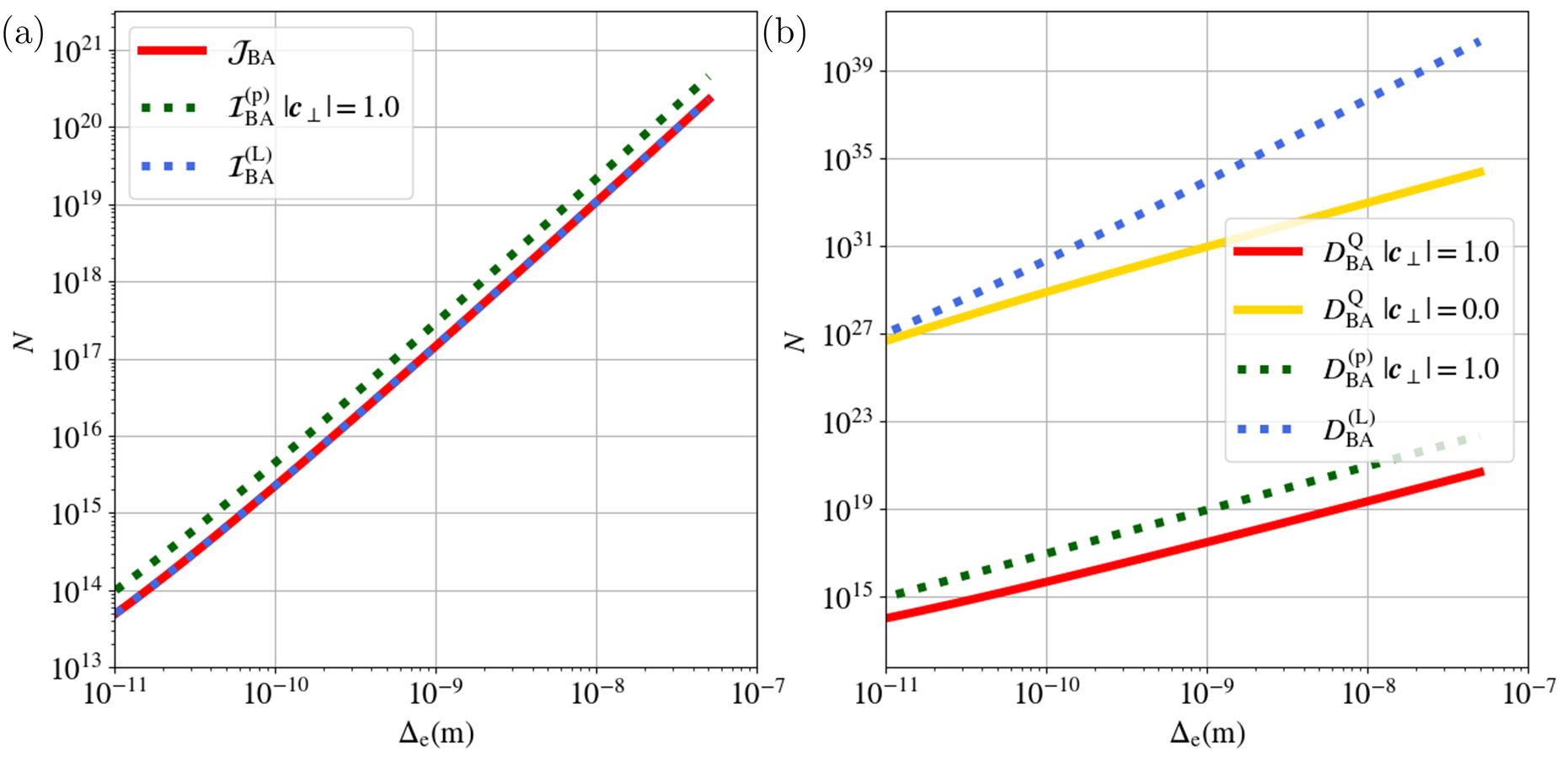}
    \caption{Required number of electrons as function of the probe size for (a) estimating the nuclear magneton ($\mu_{\rm n}=\mu_{\rm B}/1836$) with $\mathrm{SNR}=3$ and (b) discriminating its presence at $87\%$ confidence. The nuclear spin is delocalized according to a Gaussian distribution with width $\Delta_\s=1$~pm, leading to an interaction strength of $\vartheta=1.54\cdot 10^{-6}$. The classical lower bounds from momentum (dotted dark green curve) and OAM measurement (dotted blue curve) are also included.}
    \label{fig:NeNshots_nuclear}
\end{figure}

For a nuclear spin, we employ the Gaussian model with a width of $\Delta_\s=1$~pm. The nuclear magnetic moment, $\mu_{\rm n}=\mu_{\rm B}/1836$, amounts to a phase parameter $\vartheta=1.54\cdot 10^{-6}$. To ensure validity of the paraxial regime, we restrict the electron focus size to $\Delta_\e \geq 10\,\text{pm} > \Delta_\s$. In  Fig.~\ref{fig:NeNshots_nuclear} we therefore only see a monotonic increase in the required number of electrons with growing $\Delta_\e$. Even at its minimum, this number is four orders of magnitude larger than in the electron spin case. A similar behavior is observed for the number of shots required for state discrimination.

\end{document}